\documentclass[preprint,tighten,trackchanges]{aastex63}
\usepackage{color}
\usepackage[T1]{fontenc}

\begin{document}
 
\shorttitle{The study of Period change investigation of 32 CEBs in the Galactic Bulge}
\shortauthors{Hong et al.}

\title{Improved period variations of 32 contact binaries with rapidly decreasing periods \\ in the Galactic Bulge}

\author[0000-0002-8692-2588]{Kyeongsoo Hong}
\affiliation{Korea Astronomy and Space Science Institute, Daejeon 34055, Republic of Korea}
%
\author[0000-0002-5739-9804]{Jae Woo Lee}
\affiliation{Korea Astronomy and Space Science Institute, Daejeon 34055, Republic of Korea}
\affiliation{Astronomical Institute, Faculty of Mathematics and Physics, Charles University in Prague, 180 00 Praha 8, V Hole\v sovi\v ck\'ach 2, Czech Republic}

\author[0000-0000-0000-0000]{Dong-Jin Kim}
\affiliation{Korea Astronomy and Space Science Institute, Daejeon 34055, Republic of Korea}

\author[0000-0001-9339-4456]{Jang-Ho Park}
\affiliation{Korea Astronomy and Space Science Institute, Daejeon 34055, Republic of Korea}
\author[0000-0002-6687-6318]{Hye-Young Kim}
\affiliation{Department of Astronomy and Space Science, Chungbuk National University, Cheongju 28644, Republic of Korea}
\author[0000-0003-1916-9976]{Pakakaew Rittipruk}
\affiliation{National Astronomical Research Institute of Thailand, Chiang Mai 50200, Thailand}

\author[0000-0002-2641-9964]{Cheongho Han}
\affiliation{Department of Physics, Chungbuk National University, Cheongju 28644, Republic of Korea}

\correspondingauthor{Kyeongsoo Hong}
\email{kshong@kasi.re.kr}

\begin{abstract}

We present detailed analyses of updated eclipse timing diagrams for 32 contact binary merger candidates in the Galactic bulge. The photometric data was obtained from 2016 to 2021 using the Korea Microlensing Telescope Network (KMTNet) with the 1.6 m telescopes located at three southern sites (CTIO, SAAO, and SSO). The times of minimum lights were determined by applying the binary star model to full light curves created at half-year intervals from the observations. The orbital period variations of the binary systems were analyzed using the $O-C$ diagrams from our new timings with the others published in the literature (Hong et al.), which are based on the OGLE observations from 2001 to 2015. As results, the orbital periods and period decreasing rates of 32 binary systems were located to be in the ranges of 0.370$-$1.238 days and from $-3.0$ to $-13.1\times10^{-6}$ day yr$^{-1}$, respectively. Out of these stars, 24 systems show a combination effect of a parabola and a light travel time caused by a third body and their outer orbital periods are in the range of 9.1$-$26.5 yr, respectively. We propose that all of our merger candidates need additional monitoring observations to study a luminous red nova (LRN) progenitor. 
 
\end{abstract}

 
\section{Introduction}

The merger of low-mass contact eclipsing binaries (CEBs) with a rapidly decaying orbital period is associated with the transient events of luminous red novae (LRNe), which are located between classical novae and supernovae \citep{Soker+2003, Kulkarni+2007, Tylenda+2011, Ivanova+2013, Pejcha2014, Pejcha+2016, Pejcha+2017, Pastorello+2019}. The mechanisms of merger can be triggered by the Darwin instability \citep{Darwin1879}, which occurs when the spin angular momentum of the system is more than a third of its orbital angular momentum \citep[$J_{\rm orb} \leqslant 3J_{\rm spin}$,][]{Hut1980, Rasio1995, Li+2006}. This happens when the mass of the secondary component and the mass ratio are extremely small \citep{Soker+2006, Wadhwa+2021} and/or possibly in combination with other effects such as mass and angular momentum loss from the vicinity of the $L_2$ point \citep{Rasio1995, Stepien+2012, Pejcha+2016}.

Recently, \citet{Howitt+2020} summarized the plateau period and distance estimates for the recognized 13 LRNe. Among them, LRN V1309 Sco is the only known confirmed merger event, based on archive data from the OGLE by \citet{Tylenda+2011}. He reported that the precursor of the LRN was a W UMa-type contact binary with an orbital period of 1.44 days and that it showed an exponential period decay before the eruption. \citet{Nandez+2014} presented that the initial mass of V1309 Sco with a mass ratio of $\sim0.1$ was composed of 1.52 $M_{\odot}$ primary and 0.16 $M_{\odot}$ secondary. \citet{Ferreira+2019} suggested that the LRN V1309 Sco will evolve into a blue straggler state. Such stellar merger events occur about once every 10 yr in our galaxy \citep{Kochanek+2014}. The CEB precursors provide us an excellent opportunity for understanding stellar merging scenarios because they provide the fundamental physical parameters such as masses, radii, luminosities, etc.

The potential stellar merger candidates were explored using the methods such as identification of EBs with either a striking period decreasing rate \citep{Pietrukowicz+2017, Gazeas+2021, Hong+2022}, and extremely low-mass ratio below $\sim$ 0.1 \citep{Li+2017, Caton+2019, Wadhwa+2021, Li+2022, Liu+2023}. Especially, \citet{Hong+2022} presented the study for the orbital period variations of 14,127 CEBs in the Galactic bulge based on the OGLE observations between 2001 and 2015. In this study, we selected a total of 132 CEBs with period decreasing rates of higher than $-3\times10^{-6}$ day yr$^{-1}$ in the paper of \citet{Hong+2022}, which satisfy the necessary criterion for a precursor candidate by \citet{Molnar+2017}. The main goal of this study is to find out the potential merger candidates using new eclipse timings of the selected CEBs from the Korea Microlensing Telescope Network (KMTNet) observations between 2016 and 2021 in the Galactic bulge. In Section 2, the observational data of the KMTNet and OGLE are described. Section 3 presents the results of eclipse timing analysis for the stellar merger candidates. The summary and discussion of our results are presented in Section 4.

\section{Data from observations}

\subsection{KMTNet}
 
The photometric data of the selected CEBs between 2016 and 2021 in the Galactic bulge were collected from the KMTNet data archive. The KMTNet observations were monitored with an exposure time of 2 minutes using the 1.6 m wide field optical telescopes at three sites; the Siding Spring Observatory in Australia (KMTA), Cerro Tololo Interamerican Observatory in Chile (KMTC), and the South African Astronomical Observatory in South Africa (KMTS). Each CCD mosaic camera of the telescopes consisted of four 9k $\times$ 9k chips with a 4 deg$^{2}$ field of view. Twenty-seven fields of the KMTNet were covered to be nearly 96 deg$^{2}$ and they observed with cadences from $\Gamma=2$ hr$^{-1}$ to $\Gamma=8$ hr$^{-1}$ \citep[][Figure 12]{Kim+2018}. In this study, we used the light curves in $I$ band because most KMTNet observations are performed in $I$ band. All preprocessing and photometry were performed though the KMTNet pipeline \citep[cf.][]{Kim+2016, Kim+2018}, which used the difference image analysis \citep[DIA,][]{Alard+1998}. 

\subsection{OGLE}

The OGLE-III\&IV observations were carried out between 2001 and 2015 toward the Galactic bulge using the 1.3 m telescope located at the Las Campanas Observatory in Chile \citep{Udalski+2008, Udalski+2015}. The sky coverages of the OGLE-III (2001$-$2009) and OGLE-IV (2010$-$2015) were about 92 deg$^{2}$ and 182 deg$^{2}$ of the Galactic bulge, respectively. Most of the observations were taken in $I$ band, and some $V$ band images were acquired for the color measurement.

\section{Eclipse timing Variations}
 
The 132 CEBs with a high period decreasing rate of  $\geq-3\times10^{-6}$ day yr$^{-1}$ were selected from the paper of \citet{Hong+2022}, who presented the period studies of 14,127 CEBs in the Galactic bulge. From them, a total of 111 CEBs with small scatters and sufficient data points in the KMTNet observations were selected to determine their eclipse timings. We examined whether the period changes of these systems could be represented by a downward parabola and/or a light-travel-time (LTT) effect using the additional eclipse timings. The downward parabolic variation can be explained by a mass transfer from the more massive to the less massive star and/or angular momentum loss (AML) by a magnetic stellar wind. The LTT effect can be caused by an additional companion in the system. The OGLE and KMTNet observations with a time span of about 20 years allowed us to determine the period change rates ($\dot{P}$) for the selected CEBs through the observed minus computed ($O-C$) diagram, which is an efficient tool for determining subtle changes in periodic astrophysical phenomena \citep{Rovithis-Livaniou2020}. For this, we performed the approach described below.

\begin{verse}
1. It is difficult to obtain the times of minimum light by the method of \citet{Kwee+1956} from the KMTNet survey data with cadences of $\Gamma=2-8$ hr$^{-1}$. Therefore, we used the determination method of eclipse timings for the selected CEBs using the light curve synthesis method \citep[see][]{Hong+2019, Hong+2022}. In order to obtain the binary parameters, the light curves in $I$ band formed by the OGLE and KMTNet observations in 2001 to 2021 were analyzed using the Wilson-Devinney differential correction code \citep[][hereafter WD]{Wilson+1971, VanHamme+2007}. In the WD runs, the initial binary parameters were taken from the paper of \citet{Hong+2022}. We adjusted only the orbital ephemeris parameters (the epoch $T_0$, the period $P$, and the period change rate d$P$/d$t$).
The OGLE and KMTNet observations with the model light curves for 32 CEBs are presented in Figures 1 and 2, where CEBs have been identified as merger candidates using the following procedures. The basic parameters for the 32 CEBs are listed in Table 1. The resulting period change rates (d$P$/d$t$) are presented in the last columns of Tables 2-3.
\end{verse}

\begin{verse}

2. The full light curves with an interval of half a year were formed from the KMTNet observations of 2016$-$2021, and analyzed with the binary parameters by the WD code. After testing intervals of 90, 180, and 360 days, the interval was selected by considering the information loss caused by the binning process, the increase in error and the outlier occurrence rate due to the decreasing bin size. At that point, only the orbital ephemeris parameters were adjusted. Then the primary eclipse timings and their uncertainties were determined from the epoch ($T_0$), and the error in each light curve solution calculated by the WD program, respectively. The individual eclipse timings for 8 CEBs with a parabola, and 24 CEBs with two variations, are listed in Tables 4 and 5, respectively.
\end{verse}

\begin{verse}
3. In addition to these, the times of minimum light for the selected CEBs have been collected from the paper of \citet{Hong+2022}, who used the OGLE-III\&IV archive data from 2001 to 2015. In order to obtain a mean light ephemeris, the initial epochs and periods were adopted from \citet{Hong+2022}, and we applied a linear least-square fit to all eclipse timings, as follows:
\begin{equation}
 C_1 = T_0 + PE. 
\end{equation}
Here, $T_0$ is the reference epoch, $E$ denotes the epoch number for a given cycle of the binary system, and $P$ is the orbital period. Although the selected CEBs show downward parabolic variations in their $O-C$ diagrams by \citet{Hong+2022}, additional variations in the diagrams can be displayed by adding new eclipse timings from the KMTNet. Therefore, we examined whether the $O-C$ diagrams could be explained by a quadratic ephemeris or the combination of a quadratic term and an LTT ephemeris caused by the influence of an additional component, using the following equations:
\begin{equation}
 C_2 = T_0 + PE + AE^2, 
\end{equation}
where, $A$ denotes the coefficient of secular period change, and the parameters in Equation (2) were solved by a least-squares quadratic fitting.
\begin{equation}
 C_3 = T_0 + PE + AE^2 + \tau_3. 
\end{equation}
The LTT effect $\tau_3$ in Equation (3) depends on five parameters \citep{Irwin1952, Irwin1959}: $a_{\rm AB}$sin$i_{\rm out}$, $e_{\rm out}$, $\omega_{\rm out}$, $n$, and $T_{\rm peri}$. Here, $a_{\rm AB}$ denotes the projected semi-major axis of the inner binary system, and $i_{\rm out}$, $e_{\rm out}$ and $\omega_{\rm out}$ are the inclination, the eccentricity and the longitude of the periastron of the eclipsing pair around the barycentre of the three-body system, respectively. The $n$ and $T_{\rm peri}$ denote the true anomaly and the time of the periastron passage, respectively. The eight variables of the ephemeris in Equation (3) were evaluated using the Levenberg-Marquardt algorithm \citep{Press+1992}. 
To select the best-fit model, we compared two different models using the reduced $\chi^2$ of the Equation (2) and (3). As the results, a total of 32 CEBs with a period decreasing rate higher than $-3\times10^{-6}$ day yr$^{-1}$ were identified, where the eclipse timing diagrams of 8 and 24 CEBs were explained by a parabolic variation and by a combination of a parabolic plus LTT ephemeris, respectively.
\end{verse}

\subsection{Parabolic Variation}

The eclipse timing diagrams of 32 CEBs with a parabolic variation are plotted in the upper panel of Figures 3-5, where the red curves in Figure 3 and blue dashed lines in Figures 4-5 represent the quadratic terms. The resulting period decreasing rates of 32 CEBs are listed in the fourth column of Table 2 and the third column of Table 3. As one can see in Table 2, the period decreasing rates of 8 CEBs from eclipse timing analysis are in good agreement with those obtained by the WD code, while the difference between $\dot{P}_{ (O-C_3)}$ and $\dot{P}_{\rm WD}$ for about half of the 24 CEBs in Table 3 are outside their uncertainties. All of the selected CEBs were in an orbital period range of 0.370$-$1.238 days and in a period decreasing rate range between $-3.0$ and $-13.1\times10^{-6}$ day yr$^{-1}$. In our samples, the system OGLE-BLG-ECL-169991 with an orbital period of 1.176 days had the highest period decreasing rate ($\dot{P}$) of $-1.31(9)\times10^{-5}$ day yr$^{-1}$.

\subsection{Parabolic plus LTT effect}

The sinusoidal variations in the eclipse timing diagram can be explained by either a LTT effect due to an additional companion in a binary system (Irwin 1952, 1959) or a magnetic activity cycle (Applegate 1992; Lanza et al. 1998). Out of 32 CEBs, the best-fitting parameters of 24 CEBs with a quadratic term plus LTT ephemeris are listed in Table 3 together with other related quantities. The eclipse timing diagrams of 24 CEBs calculated based on Equation (3) are displayed in the top panels of Figures 4-5, where the red solid lines represent the full contribution of Equation (3). The middle and bottom panels show the LTT orbit of the third body and the residuals from the full ephemeris, respectively. The outer orbital periods of 24 CEBs are in the ranges of 9.1$-$26.5 yr.

\section{Summary and Discussion}

\citet{Hong+2022} presented a study of period variations for 14,127 CEBs in the Galactic bulge based on the OGLE observations between 2001 and 2015. Of them, a total 132 CEBs had period decreasing rates larger than $-3\times10^{-6}$ day yr$^{-1}$. The decreasing rate is the necessary criterion for a precursor candidate by \citet{Molnar+2017}. In this study, we examined the data to find suitable merger candidates using the updated eclipse timing diagrams of 111 CEBs, which had small scatters and enough data points based on the additional KMTNet observations from 2016 to 2021. The eclipse timings for the selected systems were obtained using the binary solutions from the WD code, which determined the minimum epochs without the impact of spot activity or distorted light curves \citep{Maceroni+1994, Lee+2014}. The $O-C$ diagrams of all selected CEBs showed parabolic variations. Out of these candidates, a total of 24 CEBs exhibit a quadratic term plus LTT ephemeris. The orbital periods and period decreasing rates of all selected CEBs were located within the ranges of 0.370$-$1.238 days and from $-3.0$ to $-13.1\times10^{-6}$ day yr$^{-1}$, respectively. Their outer orbital periods of 24 CEBs are in the range of 9.1$-$26.5 yr. As a results, a total of 32 potential binary merger candidates were found to satisfy the essential criterion. These systems need constant monitoring observations until the binary merging, or a revision of their period variations.

Among our candidates, OGLE-BLG-ECL-169991 with the highest period decreasing rate of $-1.31\times10^{-5}$ day yr$^{-1}$ has a relatively long orbital period of 1.176 days. Its period is similar to the LRN 1309 Sco with an orbital period of 1.43 days before eruption. \citet{Wadhwa+2021} listed the most important theoretical factors for the dynamic instability of binary systems to be the merger, such as the relationship between orbital and spin angular momentum \citep{Rasio1995}, the degree of contact \citep{Rasio+1995, Li+2006}, and angular momentum loss \citep{Stepien+2012}. The Darwin instability happens when the mass ratio is less than the theoretical limit \citep{Yang+2015}. The instability mass ratio limit for contact binaries has been predicted to be below 0.1 by several investigators \citep{Rasio1995, Li+2006, Arbutina2007, Arbutina2009, Yang+2015, Wadhwa+2021, Wadhwa+2023}, and many systems with an extremely low mass ratio have been detected, such as: V1187 Her \citep[$q\sim0.044$;][]{Caton+2019}, ZZ PsA \citep[$q\sim0.078$;][]{Wadhwa+2021}, CW Lyn \citep[$q\sim0.067$;][]{Gazeas+2021}. This makes it necessary to check the mass ratio of OGLE-BLG-ECL-169991 using the spectroscopic observations.

\acknowledgments{}
We thank the OGLE and KMTNet teams for all of the observations.
This research has made use of the KMTNet system operated by the Korea Astronomy and Space Science Institute (KASI), and the data were obtained at three host sites of CTIO in Chile, SAAO in South Africa, and SSO in Australia.
This research was supported by the KASI grant 2023-1-832-03 and by grants (2019R1A2C2085965 and 2022R1I1A1A01053320) from the National Research Foundation (NRF) of Korea.

\clearpage
\begin{figure*}[!ht]
\vspace*{0pt}
\begin{center}
\begin{tabular}{cccc}   
\vspace*{-12pt}
\hspace*{-20pt}\includegraphics[width=0.24\columnwidth]{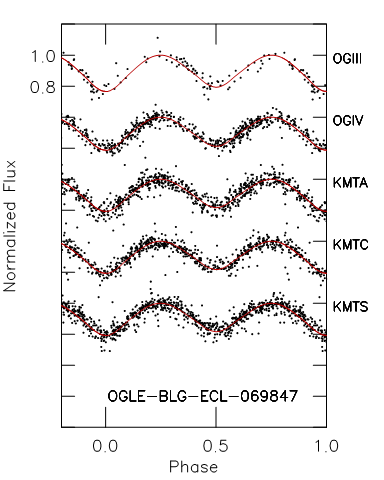} & \hspace*{-10pt}\includegraphics[width=0.24\columnwidth]{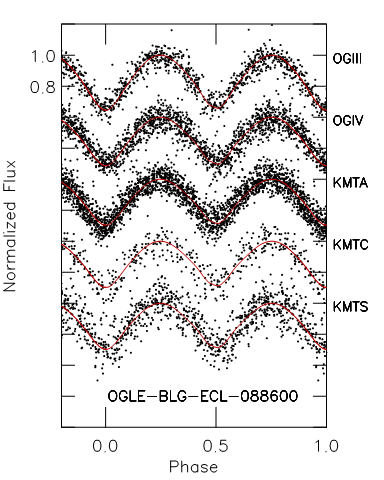} & \hspace*{-10pt}\includegraphics[width=0.24\columnwidth]{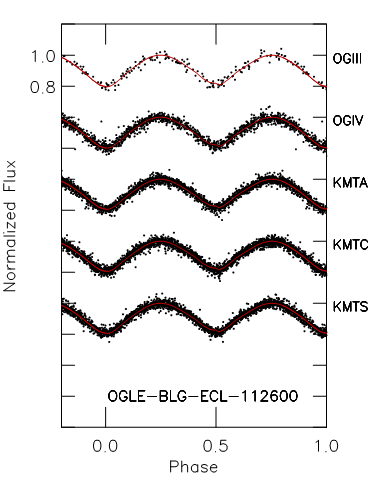}  & \hspace*{-10pt}\includegraphics[width=0.24\columnwidth]{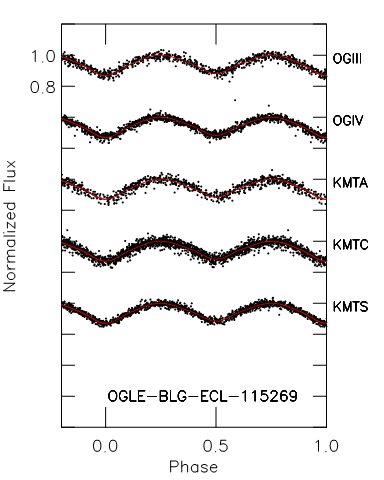} \\
\vspace*{-12pt}
\hspace*{-20pt}\includegraphics[width=0.24\columnwidth]{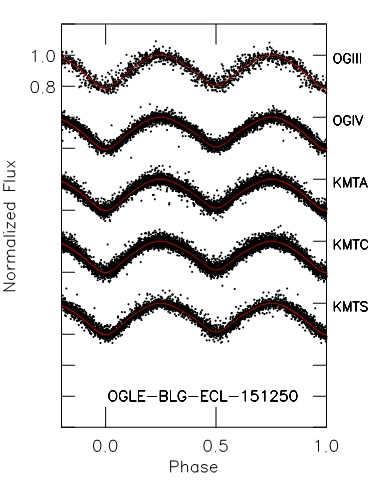} & \hspace*{-10pt}\includegraphics[width=0.24\columnwidth]{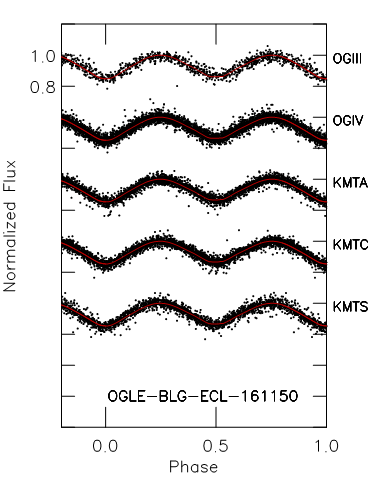} & \hspace*{-10pt}\includegraphics[width=0.24\columnwidth]{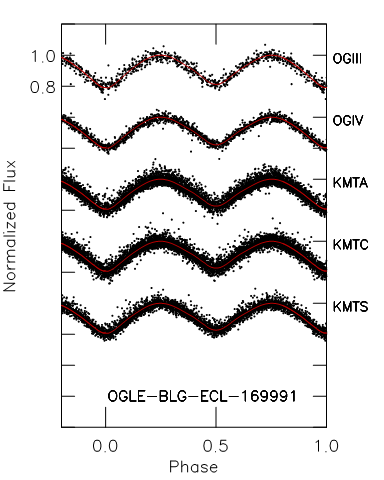}  & \hspace*{-10pt}\includegraphics[width=0.24\columnwidth]{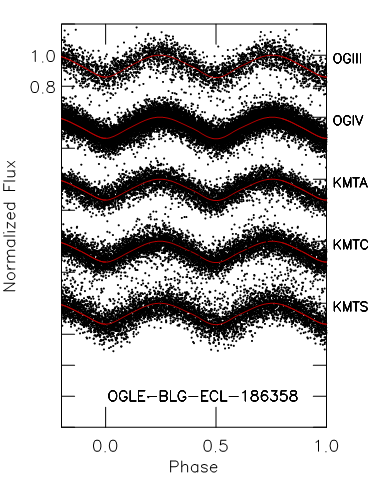} \\
\vspace*{-12pt}
\hspace*{-20pt}\includegraphics[width=0.24\columnwidth]{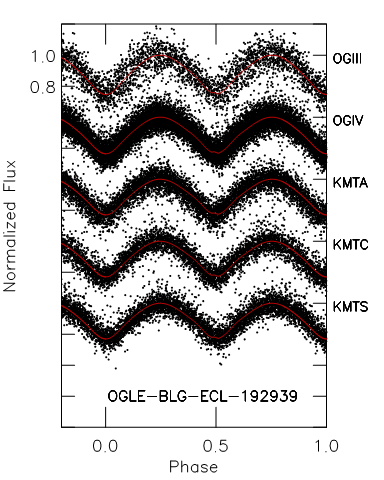} & \hspace*{-10pt}\includegraphics[width=0.24\columnwidth]{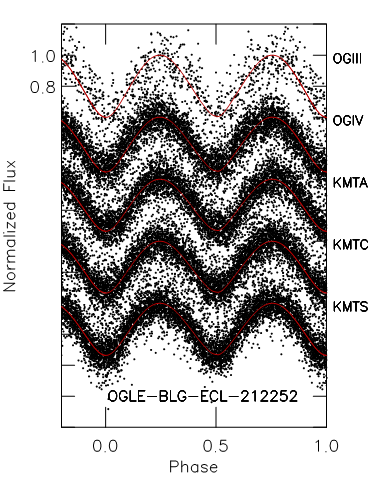} & \hspace*{-10pt}\includegraphics[width=0.24\columnwidth]{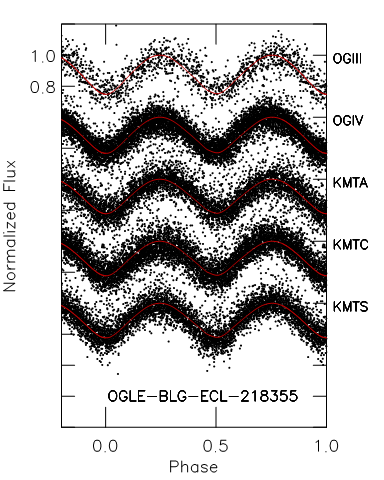}  & \hspace*{-10pt}\includegraphics[width=0.24\columnwidth]{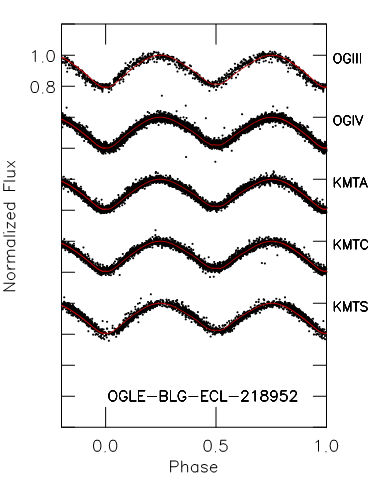} \\
\vspace*{-0pt}
\hspace*{-20pt}\includegraphics[width=0.24\columnwidth]{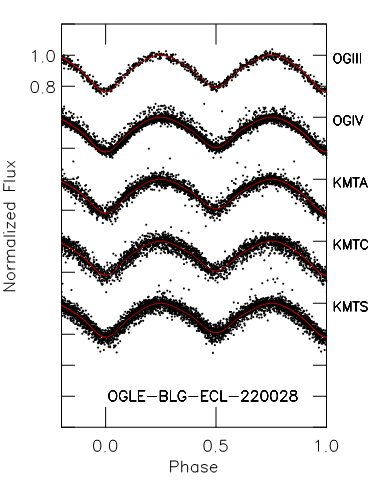} & \hspace*{-10pt}\includegraphics[width=0.24\columnwidth]{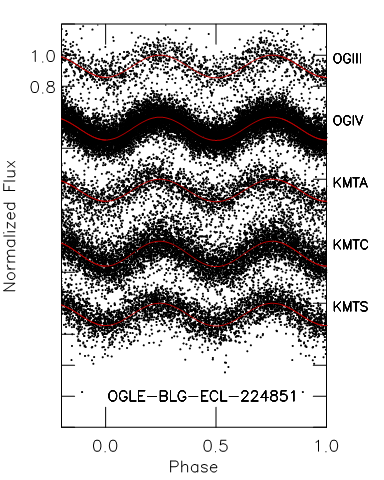} & \hspace*{-10pt}\includegraphics[width=0.24\columnwidth]{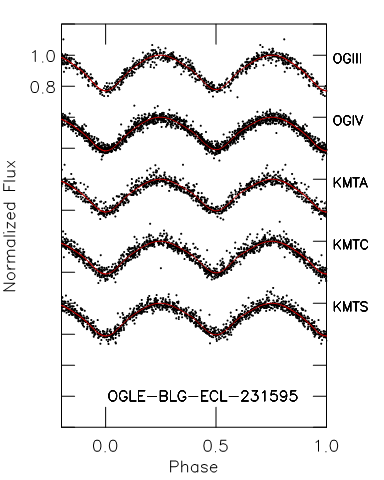}  & \hspace*{-10pt}\includegraphics[width=0.24\columnwidth]{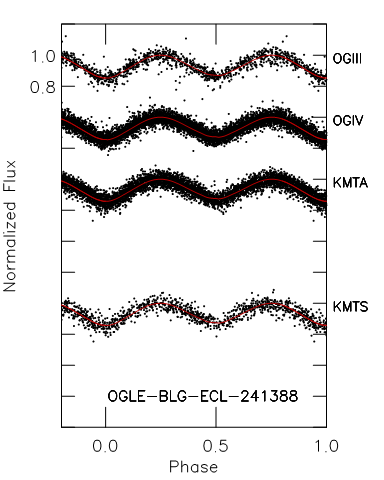} \\
\end{tabular}
\caption{$I$ light curves of 16 binary systems from the OGLE-III (OGIII), OGLE-IV (OGIV) and KMTNet surveys. The points and solid lines represent the observations and the theoretical light curves from the WD code, respectively.
}
\end{center}
\vspace*{0pt}
\end{figure*}

%
%
\clearpage
\begin{figure*}[!ht]
\vspace*{0pt}
\begin{center}
\begin{tabular}{cccc}   
\vspace*{-12pt}                                                                                                                                                                                                                                                                              
\hspace*{-20pt}\includegraphics[width=0.24\columnwidth]{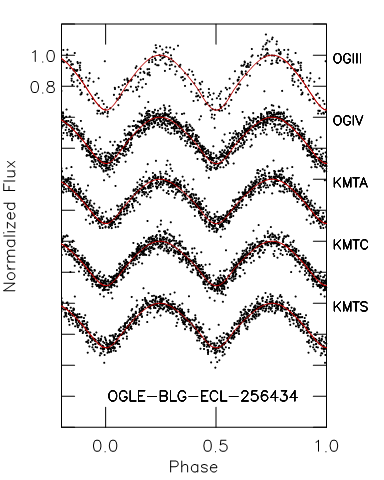} & \hspace*{-10pt}\includegraphics[width=0.24\columnwidth]{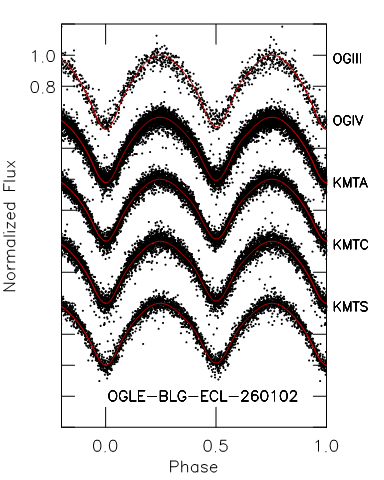} & \hspace*{-10pt}\includegraphics[width=0.24\columnwidth]{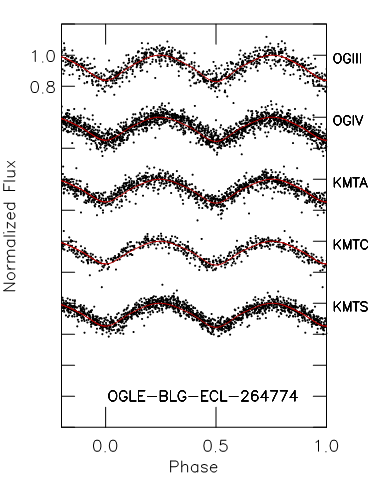}  & \hspace*{-10pt}\includegraphics[width=0.24\columnwidth]{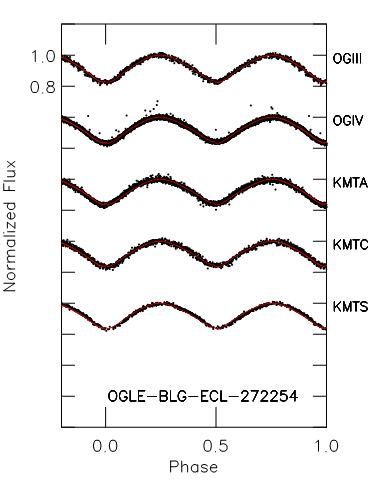} \\
\vspace*{-12pt}                                                                                                                                                                                                                                                                              
\hspace*{-20pt}\includegraphics[width=0.24\columnwidth]{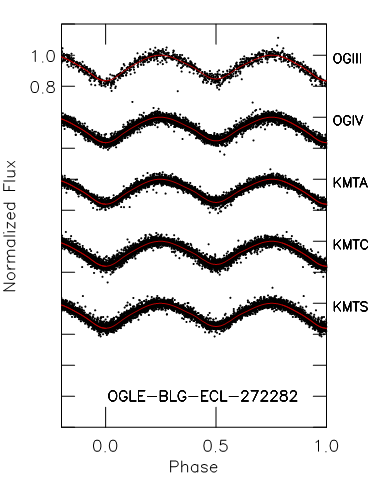} & \hspace*{-10pt}\includegraphics[width=0.24\columnwidth]{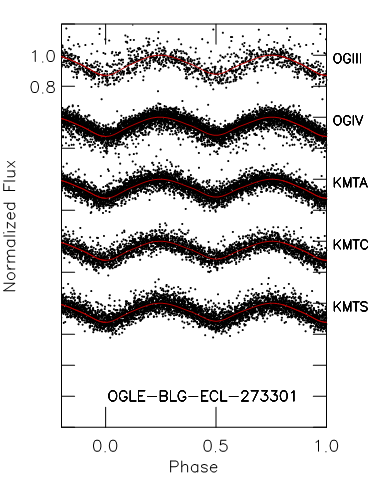} & \hspace*{-10pt}\includegraphics[width=0.24\columnwidth]{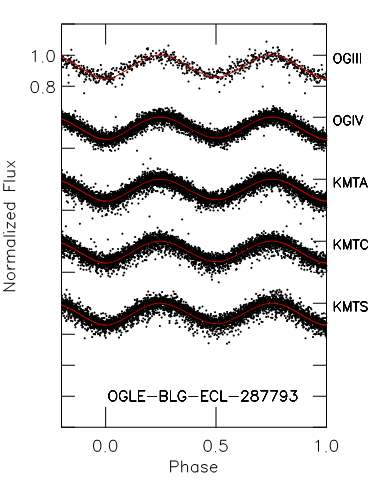}  & \hspace*{-10pt}\includegraphics[width=0.24\columnwidth]{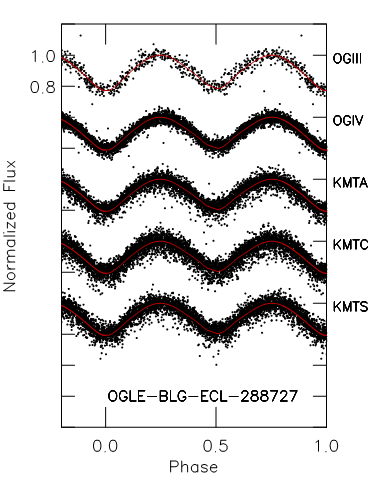} \\
\vspace*{-12pt}                                                                                                                                                                                                                                                                              
\hspace*{-20pt}\includegraphics[width=0.24\columnwidth]{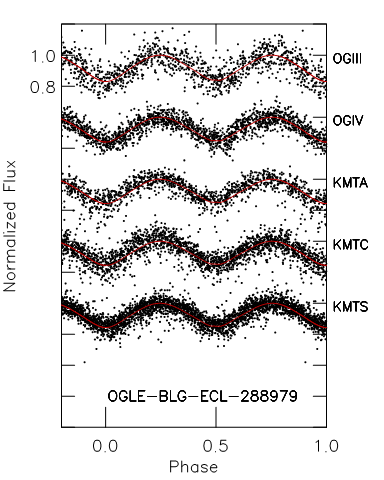} & \hspace*{-10pt}\includegraphics[width=0.24\columnwidth]{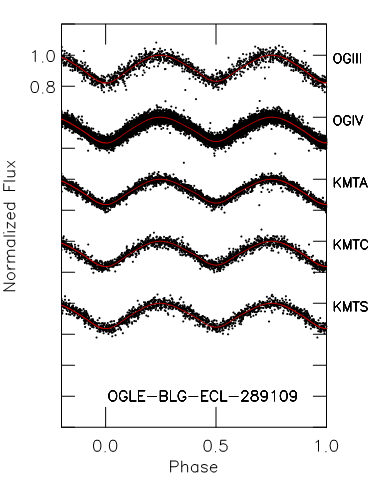} & \hspace*{-10pt}\includegraphics[width=0.24\columnwidth]{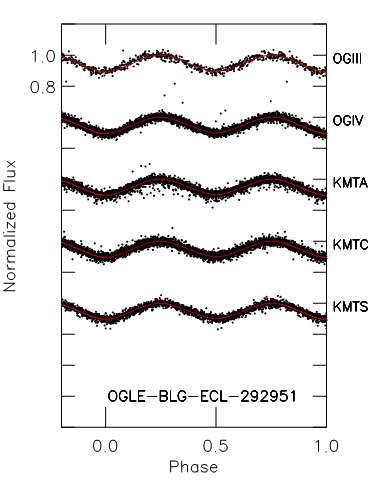}  & \hspace*{-10pt}\includegraphics[width=0.24\columnwidth]{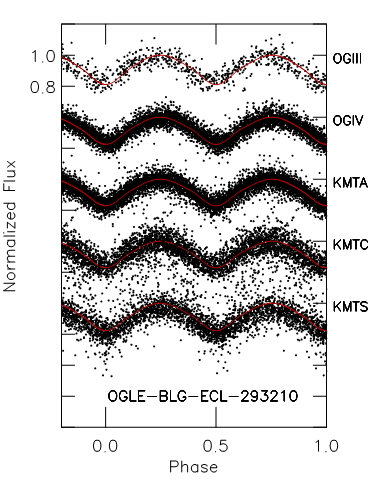} \\
\vspace*{-0pt}                                                                                                                                                                                                                                                                               
\hspace*{-20pt}\includegraphics[width=0.24\columnwidth]{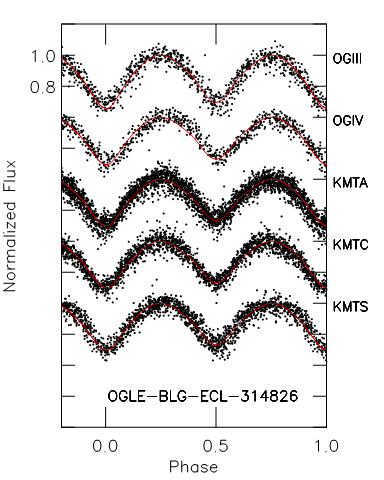} & \hspace*{-10pt}\includegraphics[width=0.24\columnwidth]{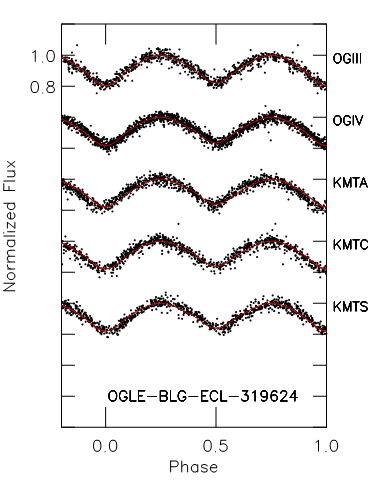} & \hspace*{-10pt}\includegraphics[width=0.24\columnwidth]{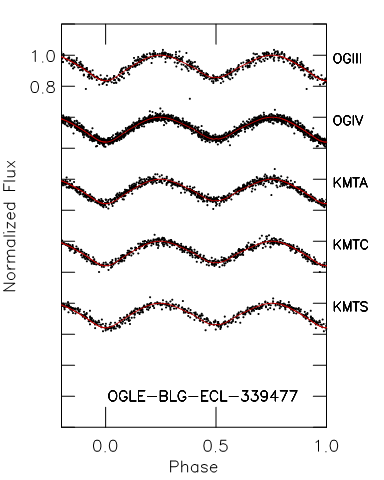}  & \hspace*{-10pt}\includegraphics[width=0.24\columnwidth]{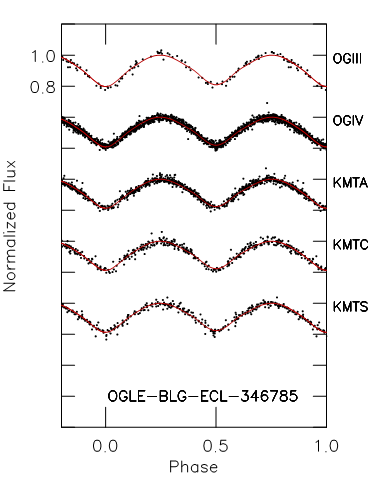} \\
\end{tabular}
\caption{Same as Figure 1, for 16 more systems.}
\end{center}
\vspace*{0pt}
\end{figure*}
%
%
%
%
%

\clearpage
\begin{figure*}[!ht]
\vspace*{0pt}
\begin{center}
\begin{tabular}{cccc}   
\hspace*{-30pt}\includegraphics[width=0.27\columnwidth]{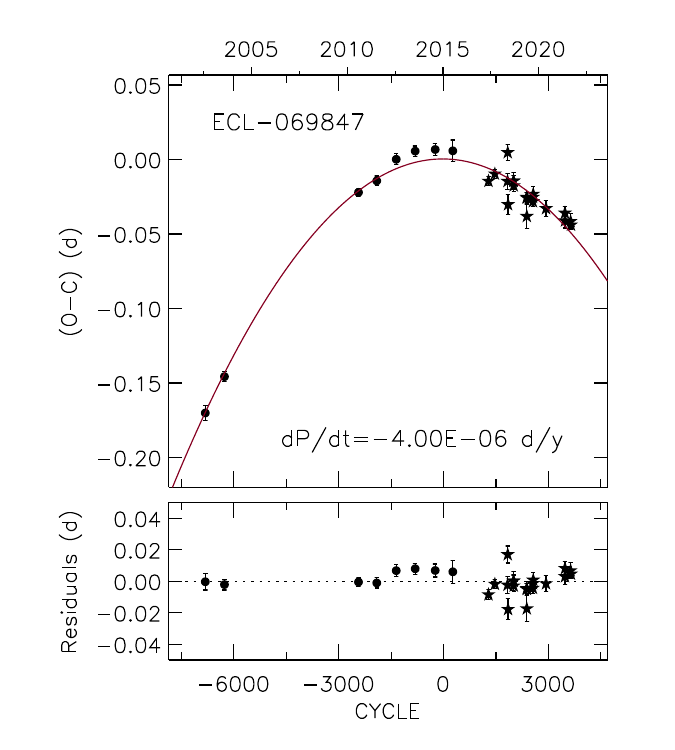} & 
\hspace*{-30pt}\includegraphics[width=0.27\columnwidth]{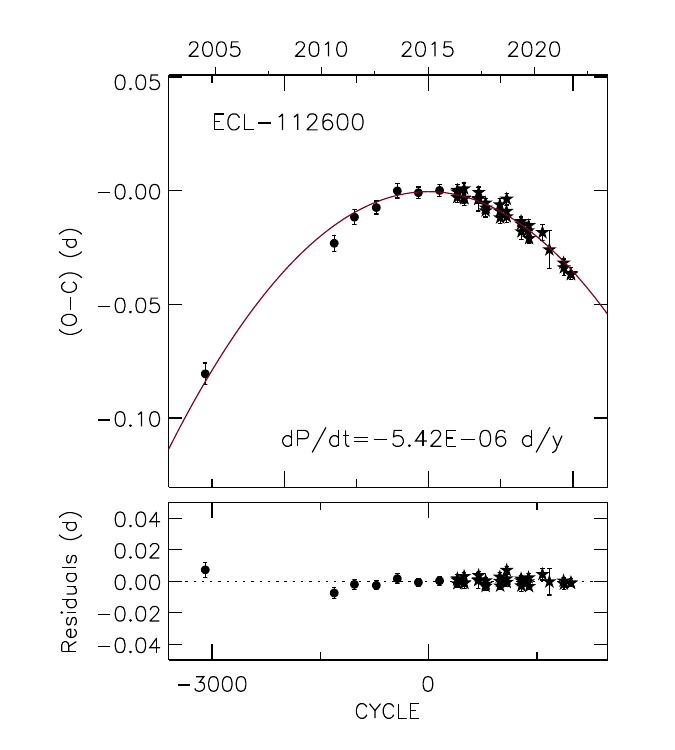} & \hspace*{-30pt}\includegraphics[width=0.27\columnwidth]{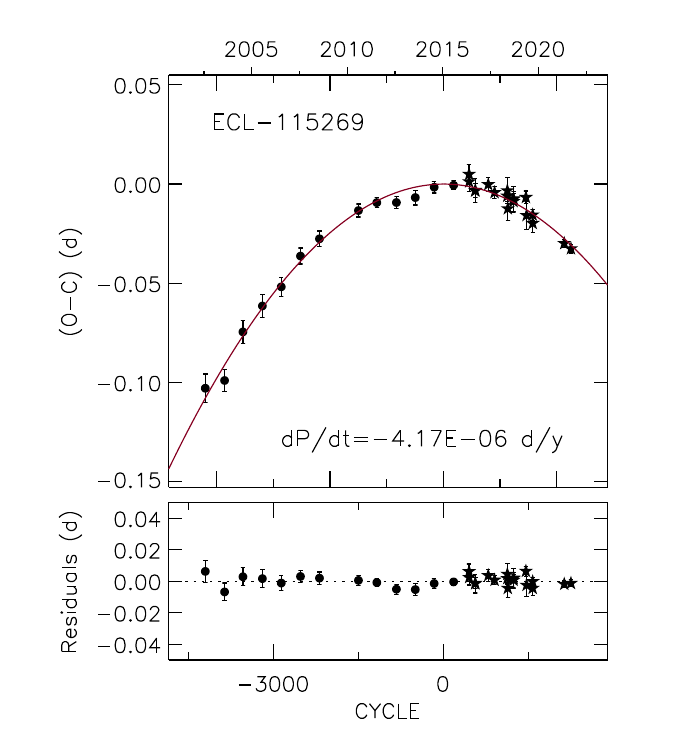}  & \hspace*{-30pt}\includegraphics[width=0.27\columnwidth]{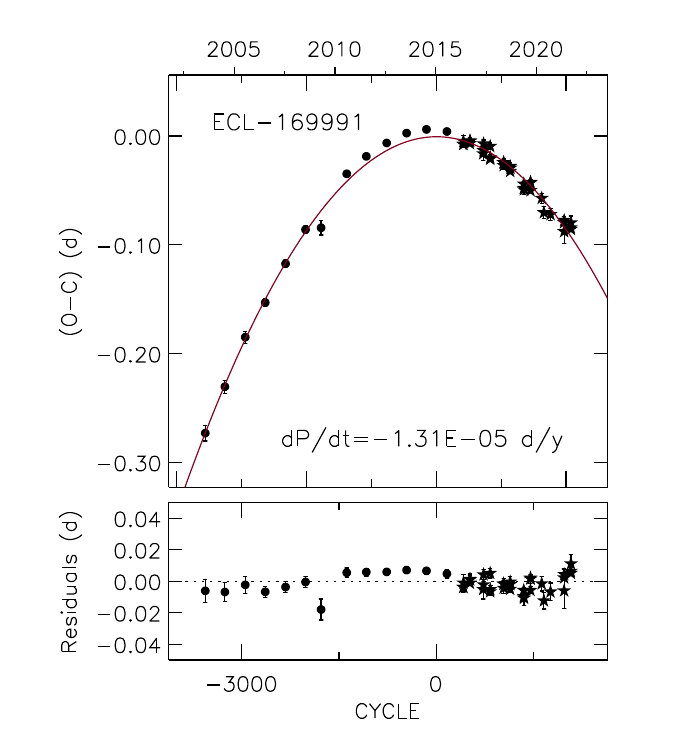} \\
\hspace*{-30pt}\includegraphics[width=0.27\columnwidth]{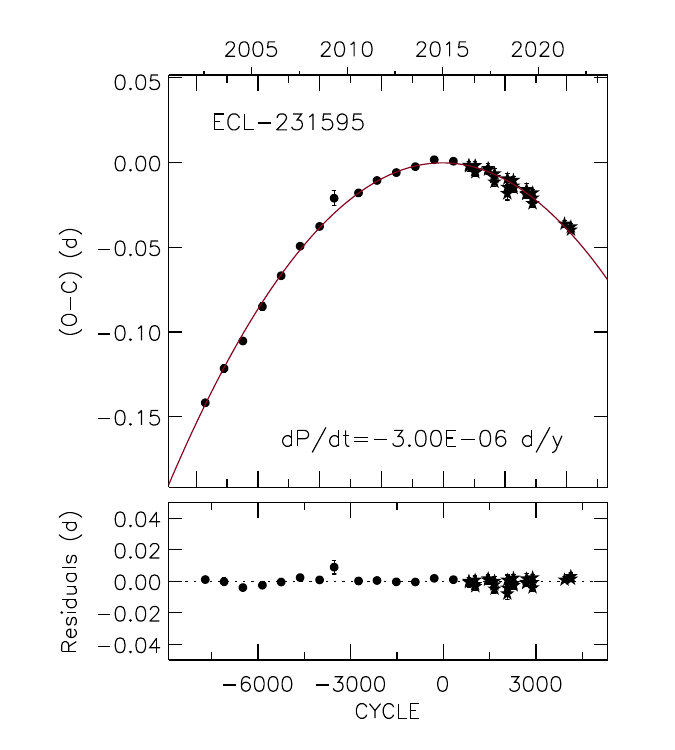} & \hspace*{-30pt}\includegraphics[width=0.27\columnwidth]{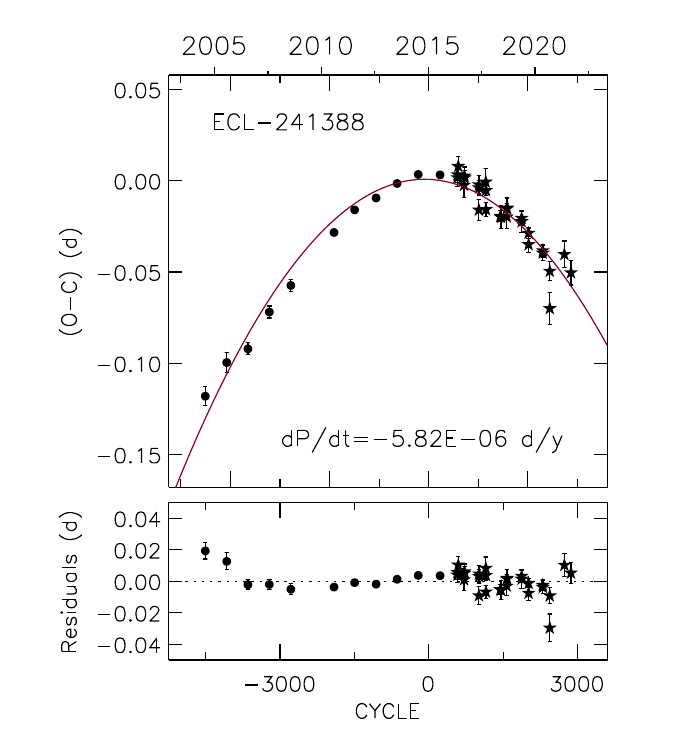} & \hspace*{-30pt}\includegraphics[width=0.27\columnwidth]{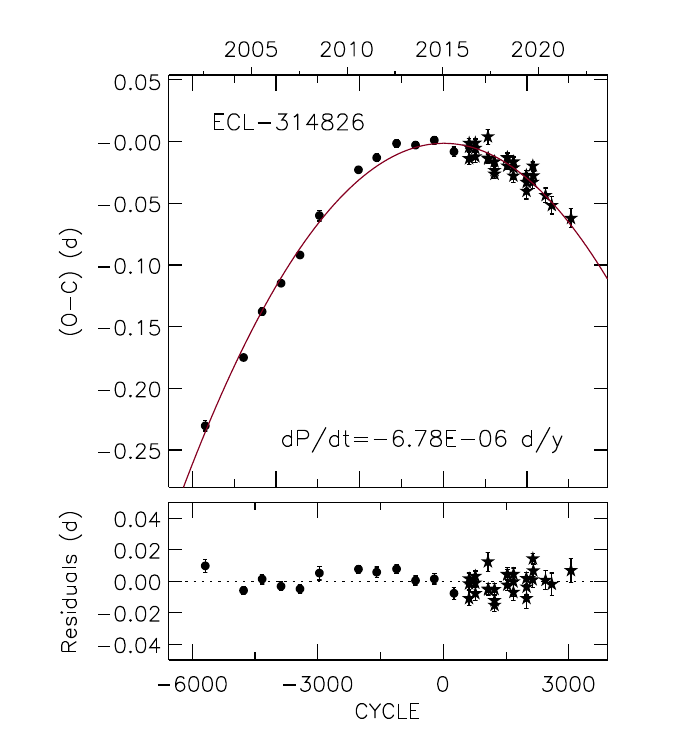}  & \hspace*{-30pt}\includegraphics[width=0.27\columnwidth]{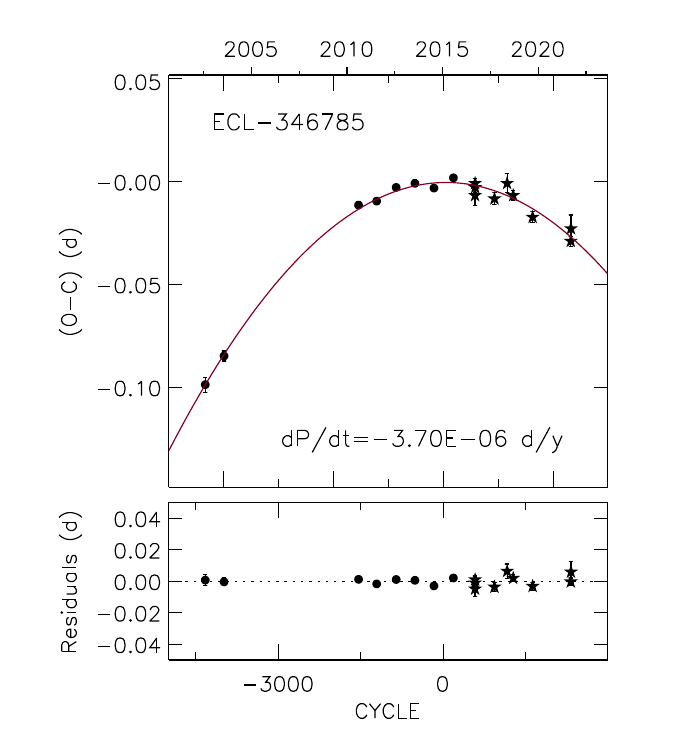} \\
\end{tabular}
\caption{Eclipse timing diagrams of eight CEBs with a parabolic variation.
}
\end{center}
\vspace*{0pt}
\end{figure*}

\clearpage
\begin{figure*}[!ht]
\vspace*{0pt}
\begin{center}
\begin{tabular}{cccc}   
\hspace*{-20pt}\includegraphics[width=0.25\columnwidth]{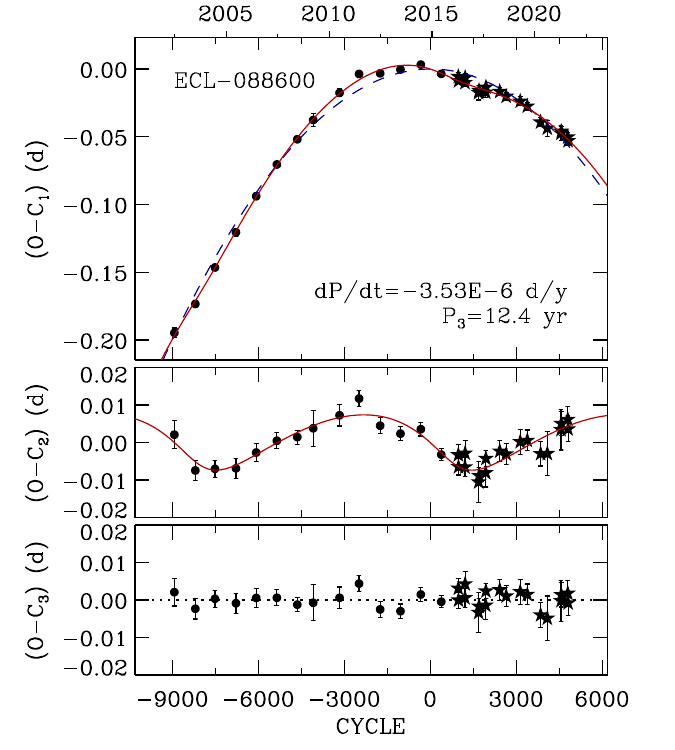} & \hspace*{-20pt}\includegraphics[width=0.25\columnwidth]{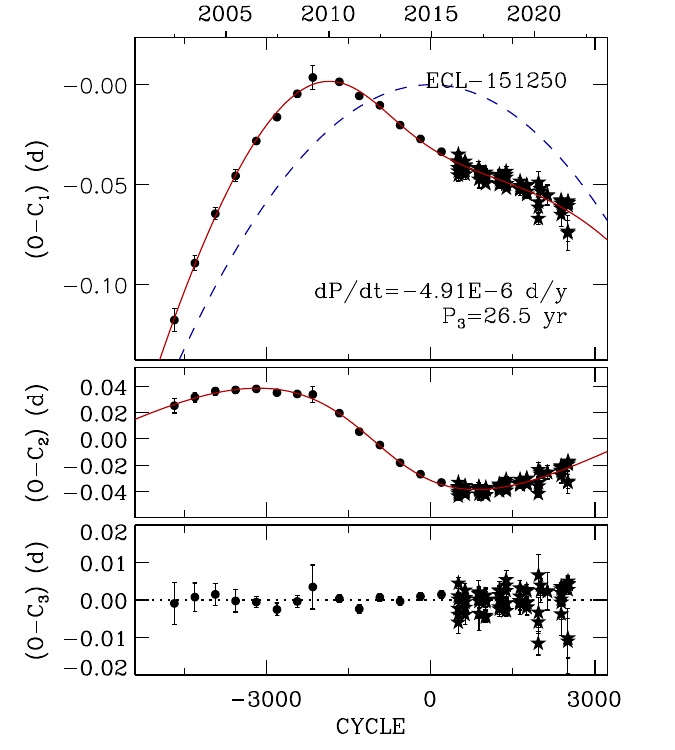} & \hspace*{-20pt}\includegraphics[width=0.25\columnwidth]{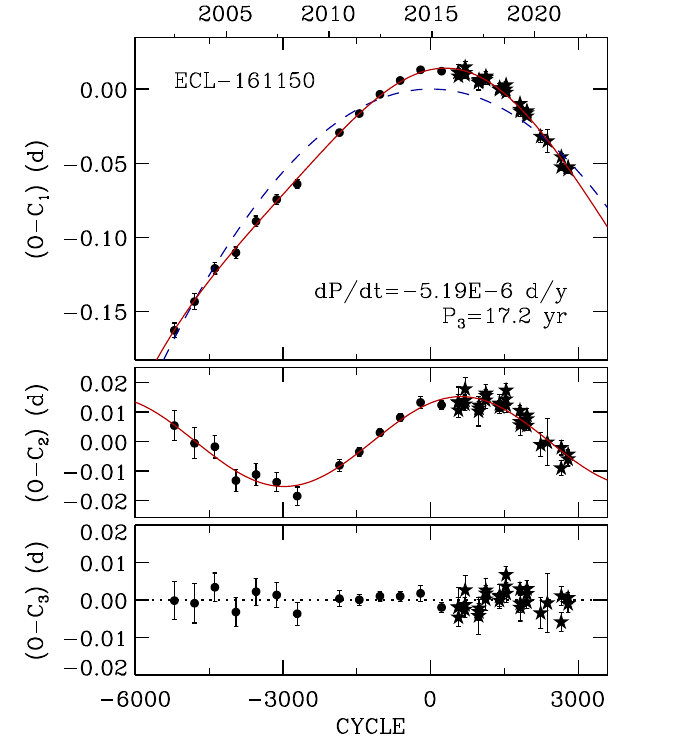}  & \hspace*{-20pt}\includegraphics[width=0.25\columnwidth]{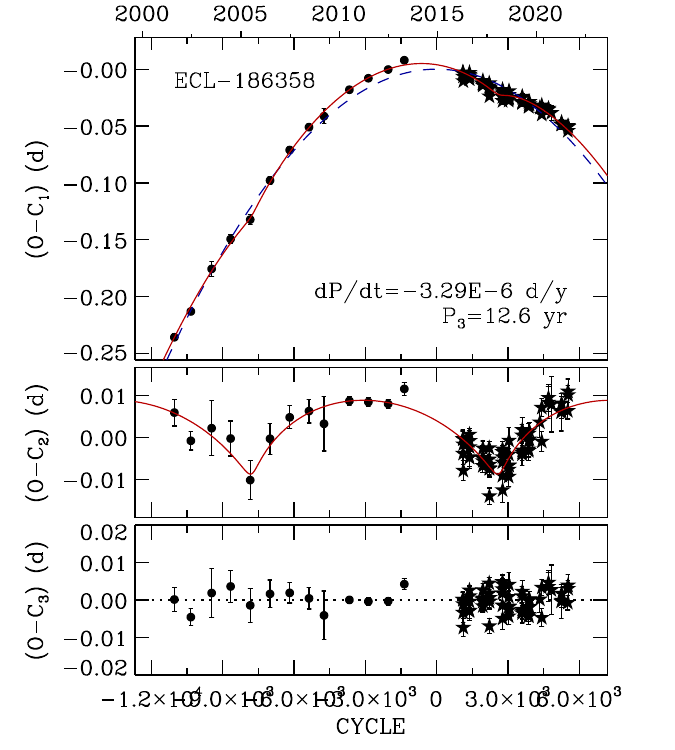} \\

\hspace*{-20pt}\includegraphics[width=0.25\columnwidth]{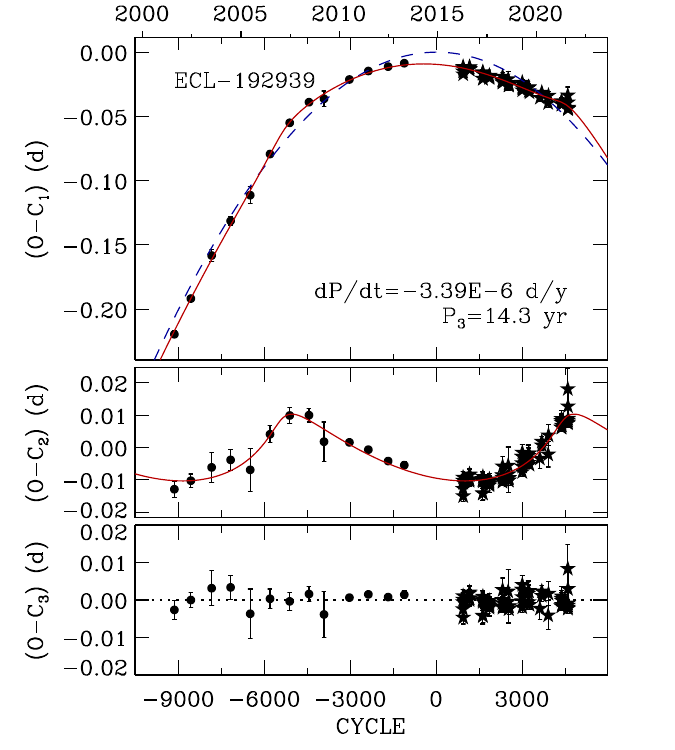} & \hspace*{-20pt}\includegraphics[width=0.25\columnwidth]{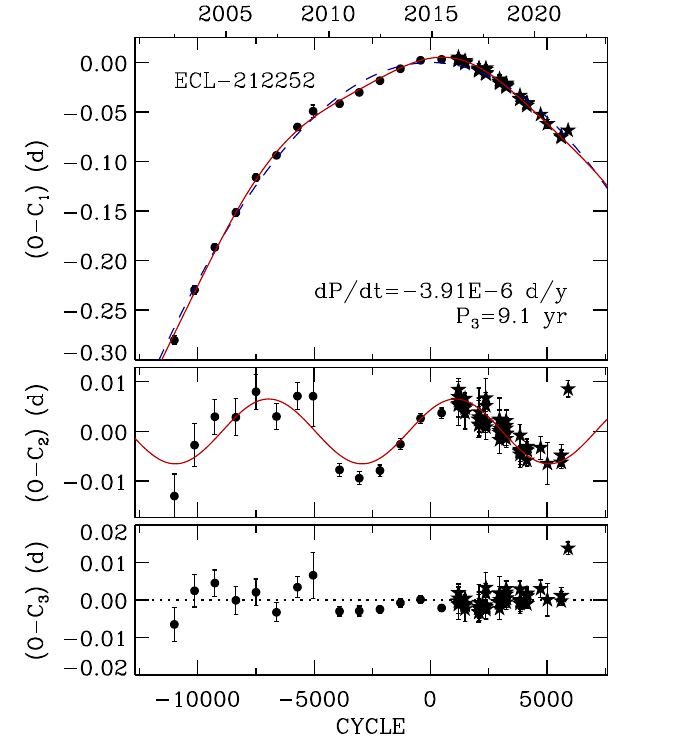} & \hspace*{-20pt}\includegraphics[width=0.25\columnwidth]{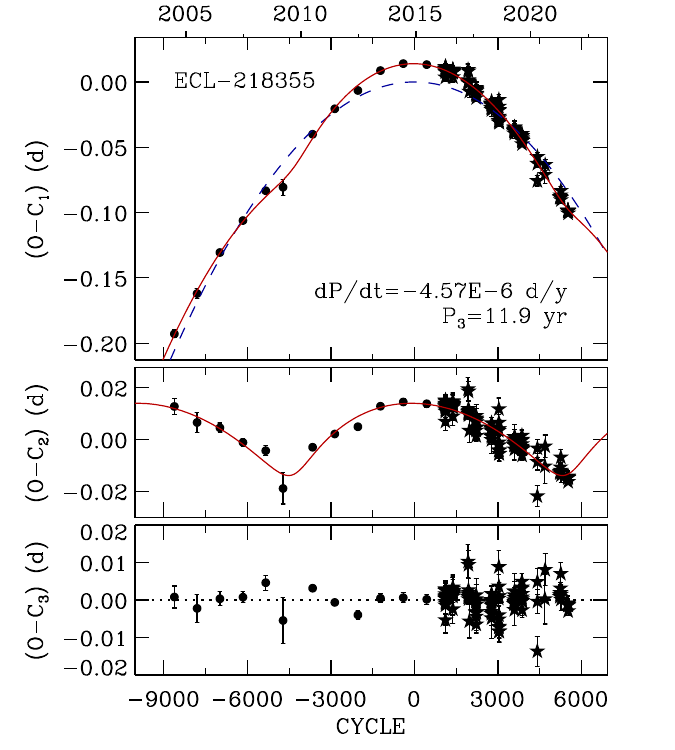}  & \hspace*{-20pt}\includegraphics[width=0.25\columnwidth]{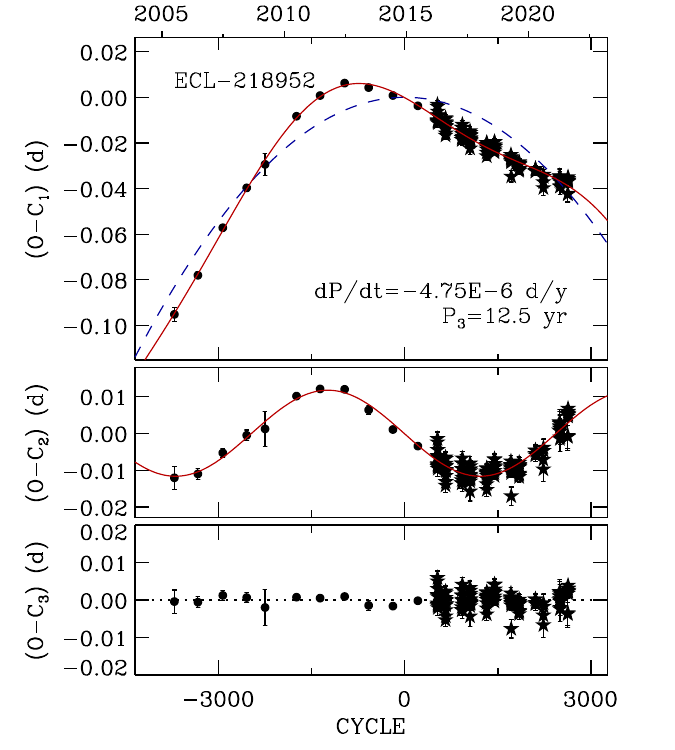} \\

\hspace*{-20pt}\includegraphics[width=0.25\columnwidth]{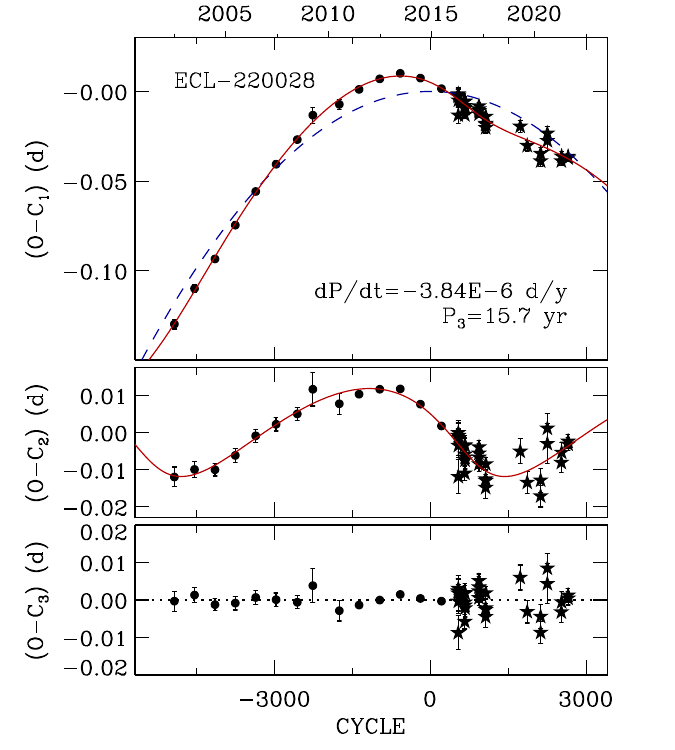} & \hspace*{-20pt}\includegraphics[width=0.25\columnwidth]{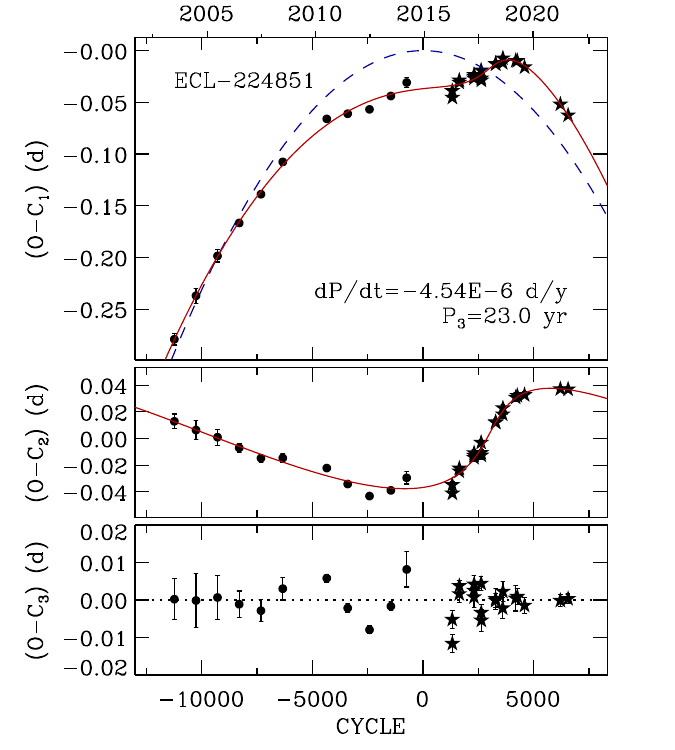} & \hspace*{-20pt}\includegraphics[width=0.25\columnwidth]{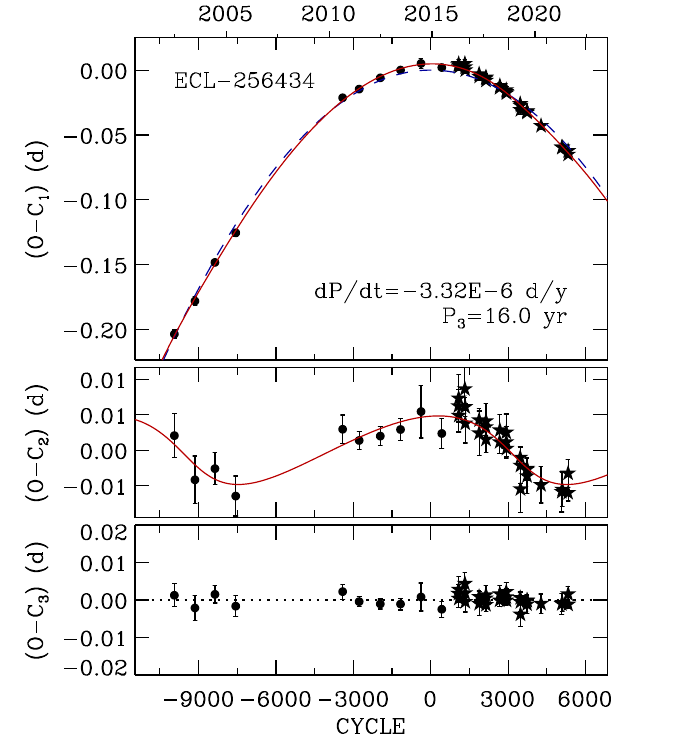}  & \hspace*{-20pt}\includegraphics[width=0.25\columnwidth]{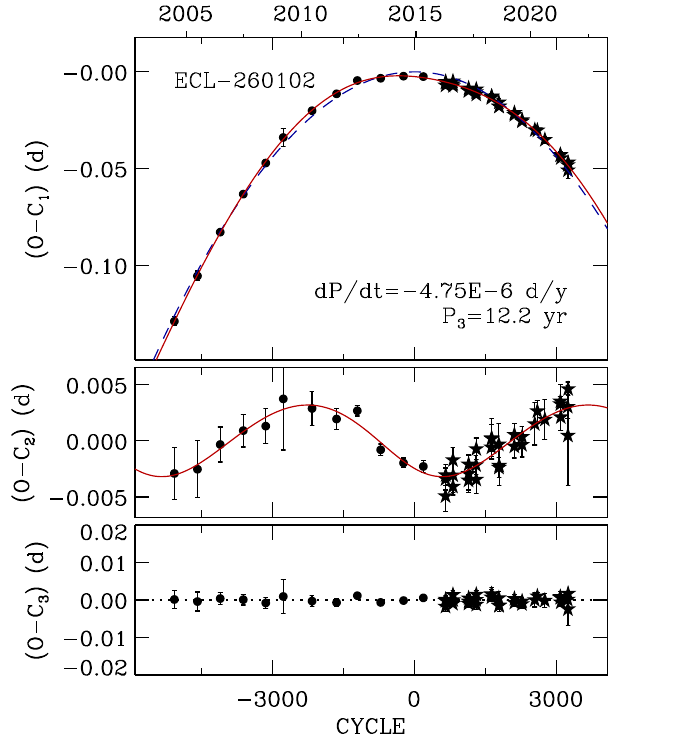} \\

\hspace*{-20pt}\includegraphics[width=0.25\columnwidth]{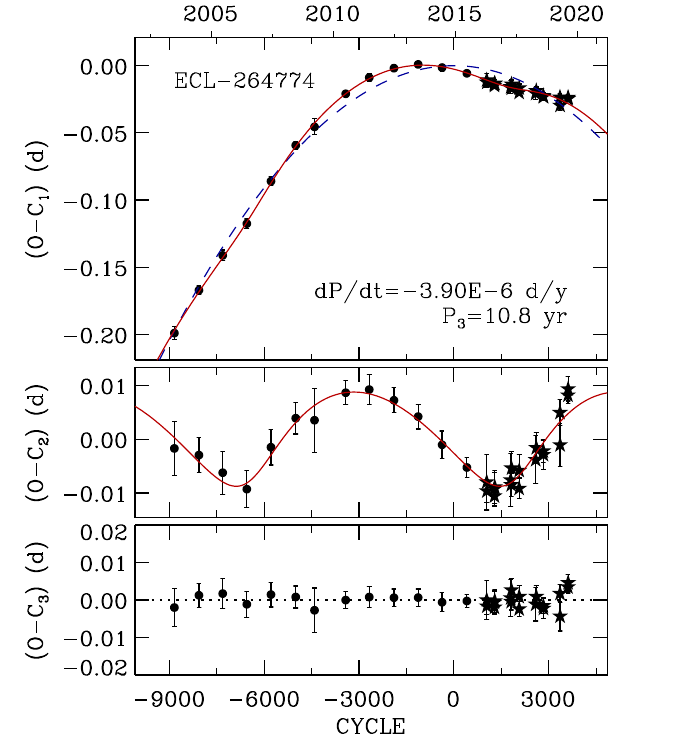} & \hspace*{-20pt}\includegraphics[width=0.25\columnwidth]{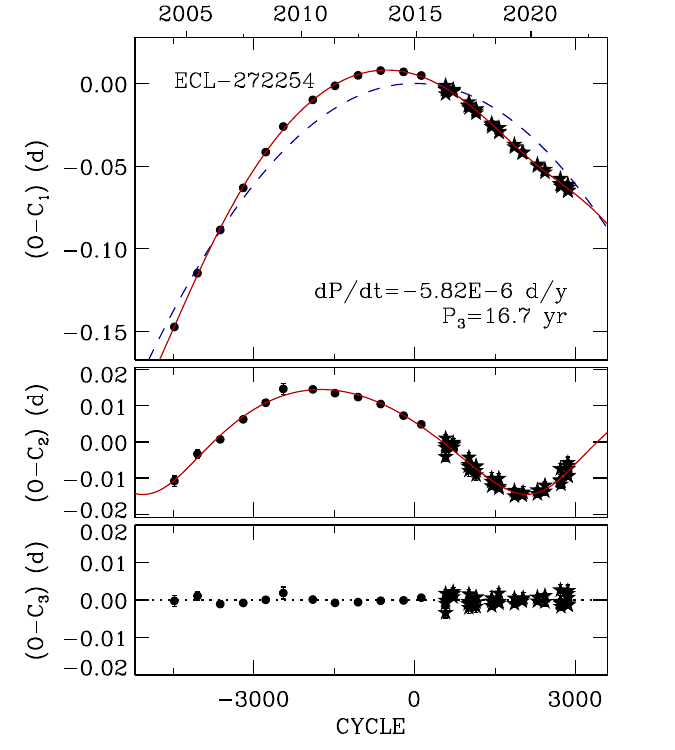} & \hspace*{-20pt}\includegraphics[width=0.25\columnwidth]{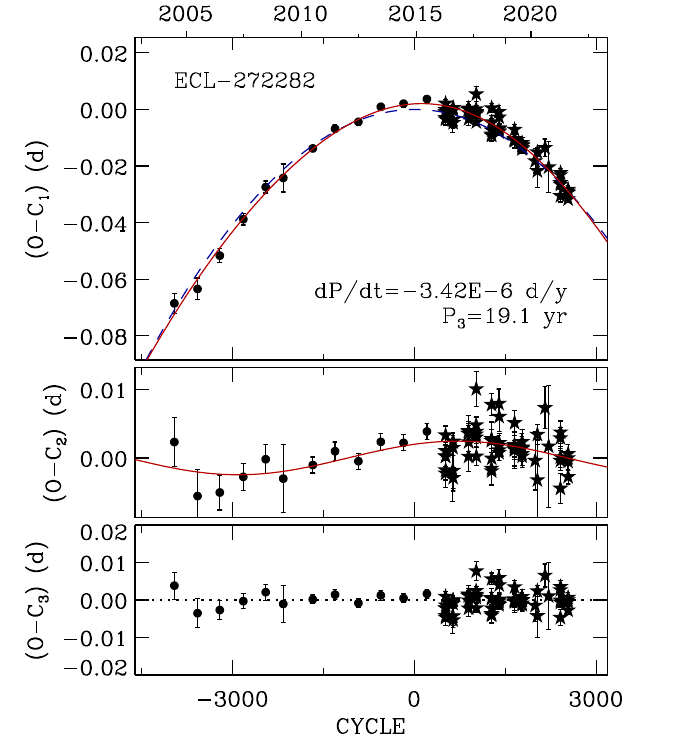}  & \hspace*{-20pt}\includegraphics[width=0.25\columnwidth]{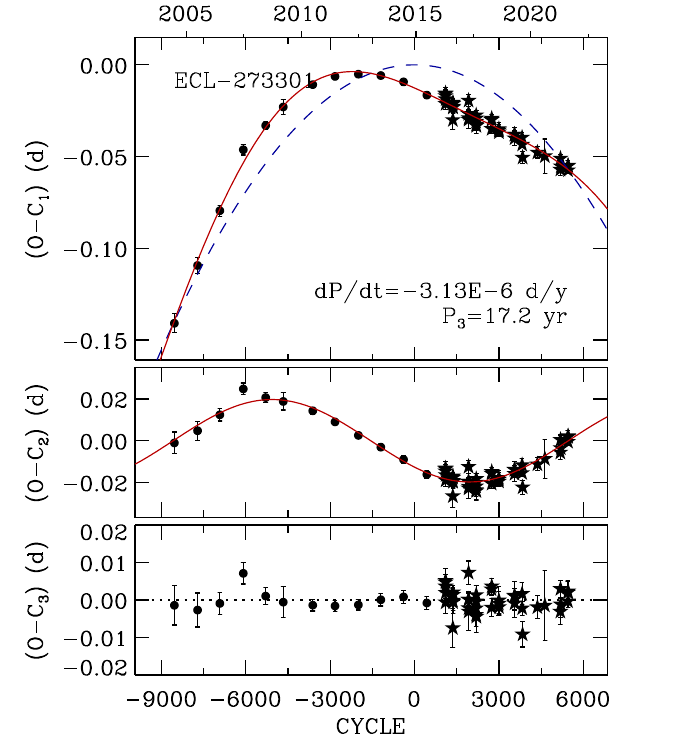} \\
\end{tabular}
\caption{Eclipse timing diagrams of 16 CEBs with a parabolic plus LTT effect.
}
\end{center}
\vspace*{0pt}
\end{figure*}

\clearpage
\begin{figure*}[!ht]
\vspace*{0pt}
\begin{center}
\begin{tabular}{cccc}   
\hspace*{-20pt}\includegraphics[width=0.25\columnwidth]{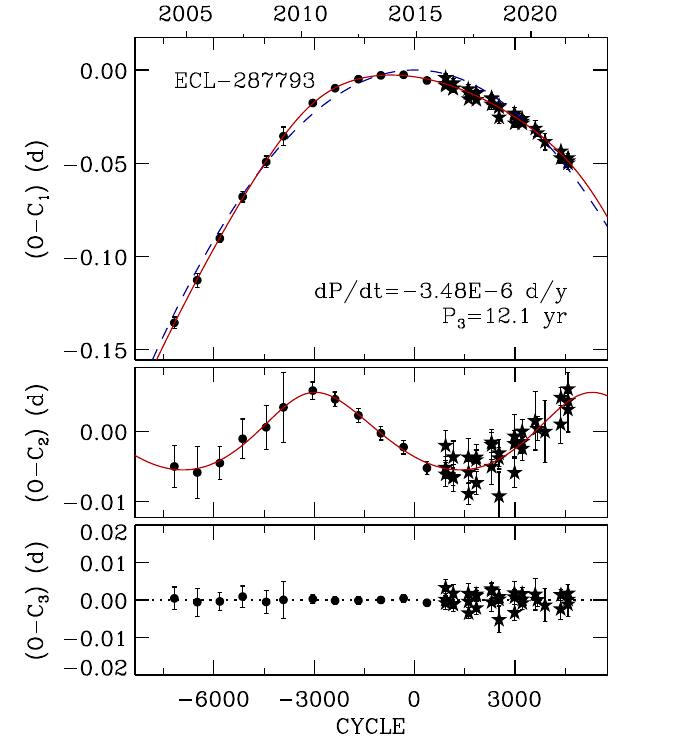} & \hspace*{-20pt}\includegraphics[width=0.25\columnwidth]{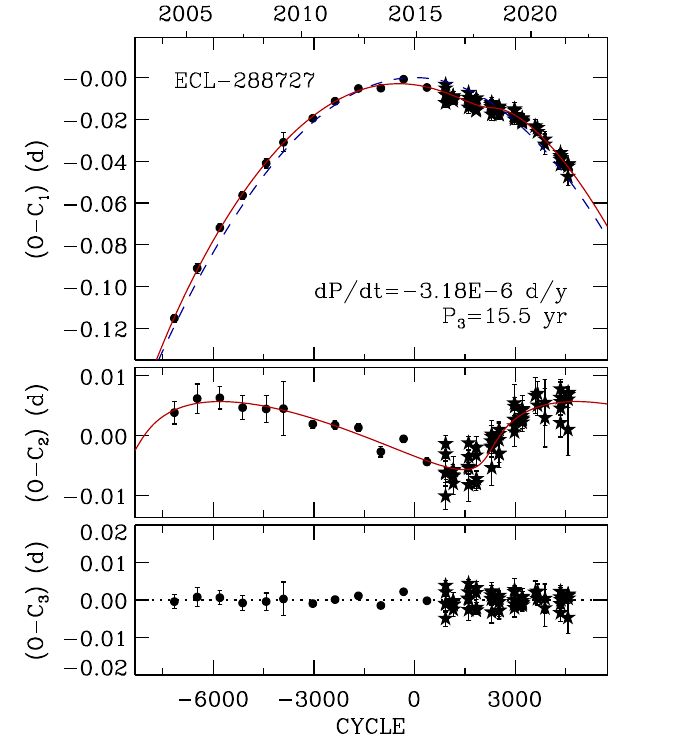} & \hspace*{-20pt}\includegraphics[width=0.25\columnwidth]{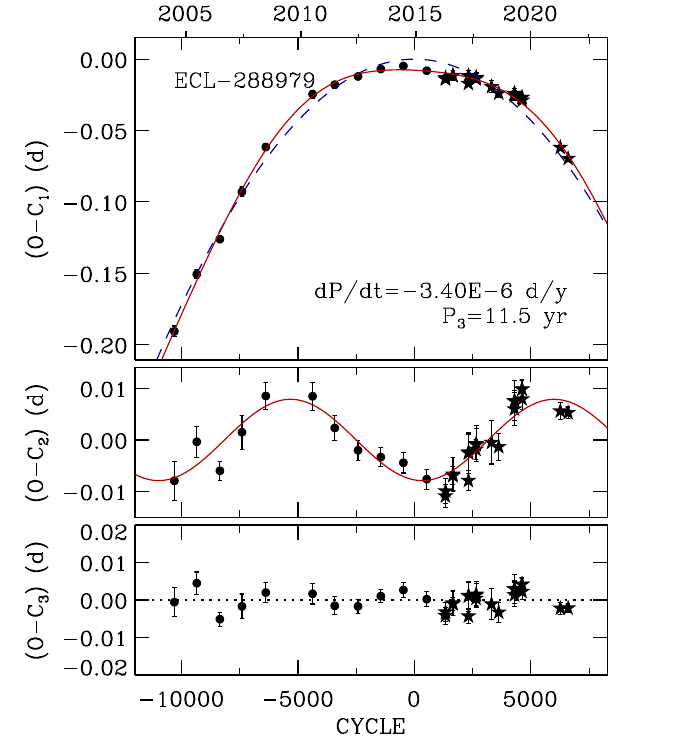}  & \hspace*{-20pt}\includegraphics[width=0.25\columnwidth]{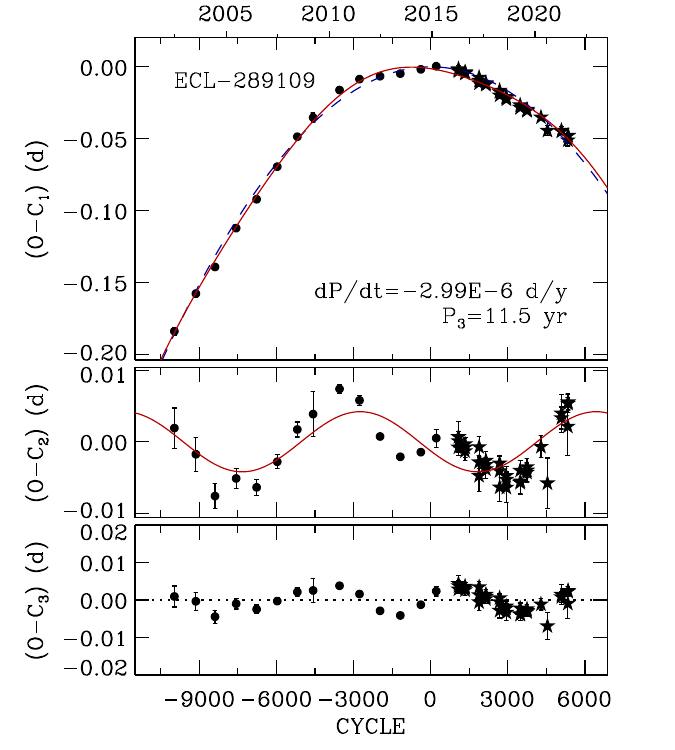} \\
\vspace*{0pt}                                                                                                                                                                                                                                                                              
\hspace*{-20pt}\includegraphics[width=0.25\columnwidth]{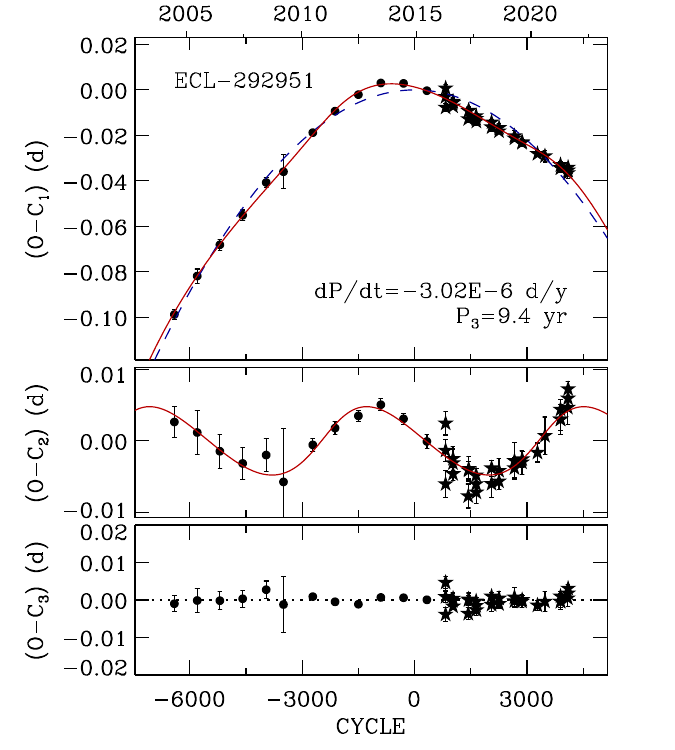} & \hspace*{-20pt}\includegraphics[width=0.25\columnwidth]{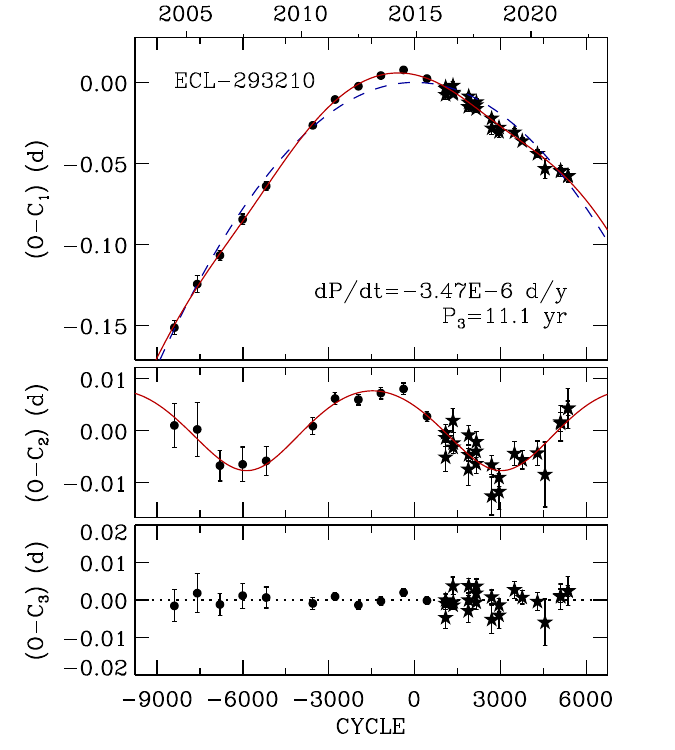} & \hspace*{-20pt}\includegraphics[width=0.25\columnwidth]{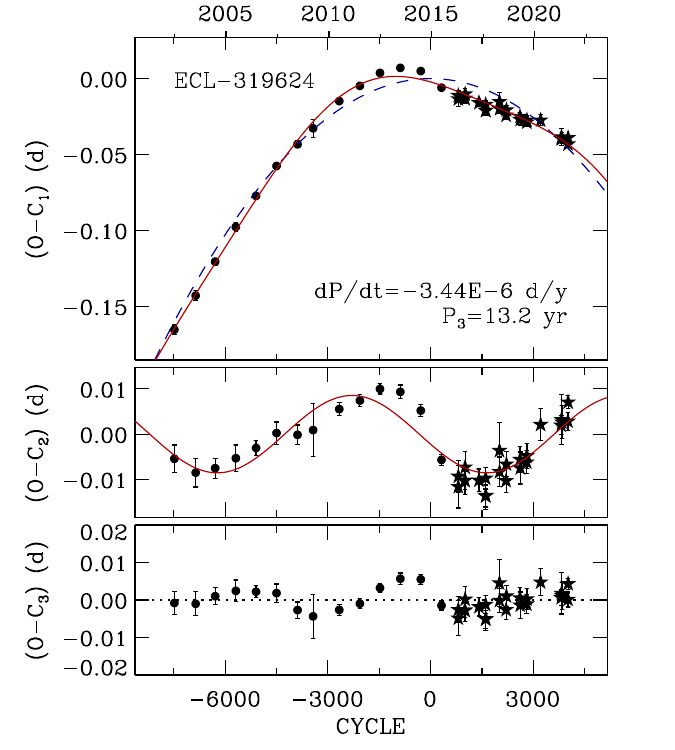}  & \hspace*{-20pt}\includegraphics[width=0.25\columnwidth]{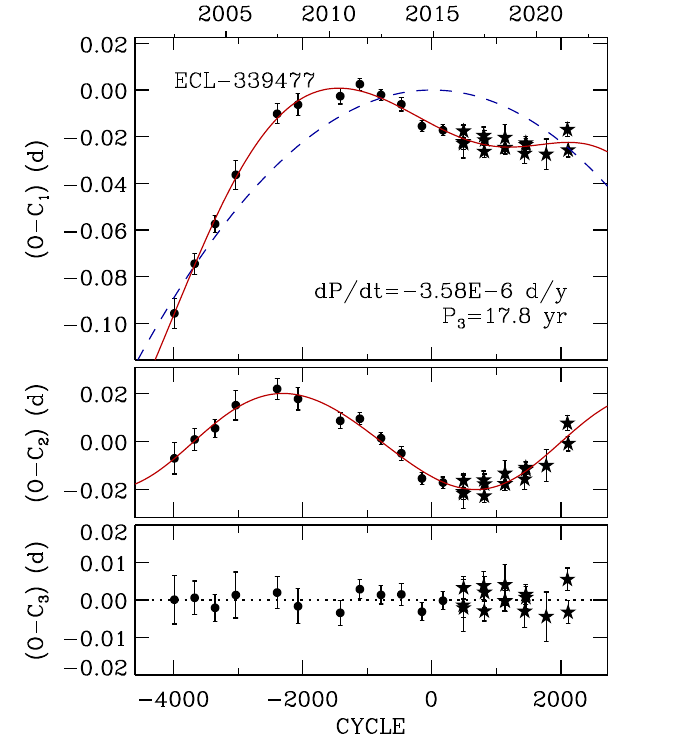} \\
\end{tabular}
\caption{Same as Figure 5, for 8 more systems.}
\end{center}
\vspace*{0pt}
\end{figure*}

\clearpage
\begin{deluxetable}{lccccccccc}
\tabletypesize{\scriptsize}
\tablewidth{0pt}
\tablecaption{Basic Parameters of 32 CEBs}
\tablehead{
\colhead{Object ID}  &\colhead{R.A.}  &\colhead{Decl.} & \colhead{$I$} &\colhead{$V-I$}  &\colhead{$P^a$}  &\colhead{$T_0^a$}  \\
\colhead{(OGLE-BLG-)}   &\colhead{(J2000)}    &\colhead{(J2000)} & \colhead{(mag)}    &\colhead{(mag)}  &\colhead{(day)}  &\colhead{(HJD+2450000)} 
}                                                                                                                                                      
\startdata                       
ECL-069847 & 17:42:03.15 & $-$22:42:46.3 & 17.910 & 1.677  &  0.6666801(26) & 2457000.4720(32) \\
ECL-088600 & 17:44:57.63 & $-$34:17:24.9 & 18.410 & 1.810  &  0.5090498(07) & 2457000.1046(10) \\
ECL-112600 & 17:48:13.97 & $-$34:12:18.7 & 17.401 & 1.753  &  1.2378942(36) & 2457000.3488(17) \\
ECL-115269 & 17:48:36.87 & $-$33:32:31.1 & 16.472 & 2.299  &  1.0832214(23) & 2457000.8871(16) \\
ECL-151250 & 17:52:36.28 & $-$30:42:53.2 & 17.127 & 2.175  &  0.9717356(09) & 2457000.7534(06) \\
ECL-161150 & 17:53:26.14 & $-$31:12:37.5 & 17.454 & 2.211  &  0.8734009(12) & 2457000.8062(09) \\
ECL-169991 & 17:54:09.00 & $-$32:05:55.0 & 17.104 & 1.965  &  1.1757013(27) & 2457000.8180(15) \\
ECL-186358 & 17:55:32.98 & $-$29:19:16.8 & 18.164 & \dots  &  0.4410626(09) & 2457000.3277(20) \\
ECL-192939 & 17:56:06.40 & $-$29:29:21.4 & 18.206 & 2.032  &  0.5318869(08) & 2457000.4773(15) \\
ECL-212252 & 17:57:44.48 & $-$31:03:40.9 & 18.765 & \dots  &  0.4132157(04) & 2457000.2266(07) \\
ECL-218355 & 17:58:16.47 & $-$28:31:11.8 & 18.344 & 2.225  &  0.4423432(05) & 2457000.2507(08) \\
ECL-218952 & 17:58:19.54 & $-$28:08:21.9 & 16.662 & 2.211  &  0.9275454(07) & 2457000.9101(05) \\
ECL-220028 & 17:58:24.67 & $-$29:57:42.0 & 16.705 & 1.492  &  0.9207411(06) & 2457000.7851(04) \\
ECL-224851 & 17:58:50.40 & $-$29:00:13.1 & 18.460 & 1.294  &  0.3723072(09) & 2457000.1468(20) \\
ECL-231595 & 17:59:26.45 & $-$31:09:49.2 & 17.386 & 1.953  &  0.5903635(06) & 2457000.2168(08) \\
ECL-241388 & 18:00:18.16 & $-$28:13:40.4 & 16.905 & 1.723  &  0.8478846(14) & 2457000.7616(11) \\
ECL-256434 & 18:01:38.90 & $-$32:15:24.3 & 18.134 & 1.451  &  0.4576949(08) & 2457000.2088(14) \\
ECL-260102 & 18:01:59.20 & $-$28:09:04.1 & 17.934 & 1.454  &  0.7524082(05) & 2457000.1329(03) \\
ECL-264774 & 18:02:25.72 & $-$30:07:51.9 & 17.861 & 1.693  &  0.4728997(09) & 2457000.3677(13) \\
ECL-272254 & 18:03:07.25 & $-$28:15:19.0 & 15.549 & 1.168  &  0.8535252(04) & 2457000.8072(02) \\
ECL-272282 & 18:03:07.43 & $-$27:34:04.2 & 17.075 & 1.845  &  0.9639203(12) & 2457000.2507(08) \\
ECL-273301 & 18:03:12.84 & $-$27:38:38.8 & 17.456 & 1.888  &  0.4475261(07) & 2457000.2124(11) \\
ECL-287793 & 18:04:32.28 & $-$26:53:29.4 & 17.208 & 1.358  &  0.5318892(06) & 2457000.2048(06) \\
ECL-288727 & 18:04:37.09 & $-$27:05:54.1 & 17.352 & 1.768  &  0.5338122(04) & 2457000.2634(04) \\
ECL-288979 & 18:04:38.47 & $-$29:58:13.5 & 17.993 & 1.328  &  0.3699377(07) & 2457000.3433(12) \\
ECL-289109 & 18:04:39.30 & $-$29:05:08.8 & 17.052 & 1.363  &  0.4565416(03) & 2457000.0349(03) \\
ECL-292951 & 18:05:00.38 & $-$26:57:52.2 & 16.669 & 1.444  &  0.5956858(05) & 2457000.4901(06) \\
ECL-293210 & 18:05:02.07 & $-$26:57:45.3 & 17.807 & 1.428  &  0.4549985(05) & 2457000.2194(07) \\
ECL-314826 & 18:07:08.24 & $-$30:22:00.3 & 17.601 & 1.630  &  0.7990017(18) & 2457000.7918(20) \\
ECL-319624 & 18:07:38.73 & $-$27:28:18.3 & 17.079 & \dots  &  0.6078540(06) & 2457000.1025(08) \\
ECL-339477 & 18:09:50.66 & $-$25:58:29.2 & 16.812 & 1.535  &  1.1381204(25) & 2457001.1215(15) \\
ECL-346785 & 18:10:40.50 & $-$25:11:59.9 & 16.351 & 1.628  &  1.0501516(14) & 2457000.8382(09) \\
\enddata                                                                                                            
 \begin{list}{}{}
 \item Note: The coordinates, $I$, and $V-I$ are taken from the OGLE-IV catalogue by Soszy\'nski et al. (2016). 
 \item $^{\rm a}$ Value taken from Hong et al. (2022).
\end{list}
\end{deluxetable}

%
 
\begin{deluxetable}{lcccc}
\tabletypesize{\scriptsize}
\tablewidth{0pt}
\tablecaption{The Secular Period Change Rates of 8 CEBs}
\tablehead{
\colhead{Object ID}  &\colhead{$P^a$}       &\colhead{$T_0^a$}   & \colhead{$\dot{P}_{\rm (O-C_2)}$}  & \colhead{$\dot{P}_{\rm WD}^b$}    \\                       
\colhead{(OGLE-BLG-)} &\colhead{(day)}       &\colhead{(HJD+2450000)}    &\colhead{(day year$^{-1}$)}    &\colhead{(day year$^{-1}$)} 
}                                         
\startdata                                                                                                                                             
ECL-069847.out   &      0.6666749(3)   &  2457000.4626(11)  &  $-4.00(10)$E$-06$  &   $-3.89(9)$E$-06$  \\
ECL-112600.out   &      1.2378877(6)   &  2457000.3483(7)   &  $-5.42(19)$E$-06$  &   $-5.40(11)$E$-06$ \\
ECL-115269.out   &      1.0832219(4)   &  2457000.8888(9)   &  $-4.17(15)$E$-06$  &   $-4.14(11)$E$-06$ \\
ECL-169991.out   &      1.1757003(3)   &  2457000.8113(6)   &  $-1.31(13)$E$-05$  &   $-1.30(9)$E$-05$ \\
ECL-231595.out   &      0.5903635(1)   &  2457000.2160(4)   &  $-3.00(4) $E$-06$  &   $-2.96(3)$E$-06$  \\
ECL-241388.out   &      0.8478777(3)   &  2457000.7575(6)   &  $-5.82(13)$E$-06$  &   $-5.73(10)$E$-06$ \\
ECL-314826.out   &      0.7990036(4)   &  2457000.7909(8)   &  $-6.78(12)$E$-06$  &   $-5.12(6)$E$-06$  \\
ECL-346785.out   &      1.0501509(4)   &  2457000.8377(6)   &  $-3.70(11)$E$-06$  &   $-3.58(8)$E$-06$  \\
\enddata                                                                                                            
\end{deluxetable}       
%
 
\begin{deluxetable}{lccccccccc}
\tabletypesize{\tiny} 
\tablewidth{0pt}
\tablecaption{The Fitted Parameters for the Parabolic $plus$ LTT ephemeris of 24 CEBs}
\tablehead{
\colhead{Object ID}  &\colhead{Period}       &\colhead{$\dot{P}_{\rm (O-C_3)}$}   & \colhead{$P_3$}      &\colhead{$\omega_{\rm out}$}  &\colhead{$e_{\rm out}$}   &\colhead{MHJD} &\colhead{$a_{\rm AB}$sin$i_{\rm out}$}   &\colhead{$f(m)$}  & \colhead{$\dot{P}_{\rm WD}$}    \\                       
\colhead{(OGLE-BLG-)}           &\colhead{(day)}    &\colhead{(day year$^{-1}$)} & \colhead{(year)}    &\colhead{(degree)}  &\colhead{}       &\colhead{(HJD+2450000)}       &\colhead{(AU)}  &\colhead{}       &\colhead{(day year$^{-1}$)}
}                                                                                                                                                      
\startdata                                                                                                         
ECL-088600 &    0.50905239(15)  &  $  -3.53(3)$E$-6 $  &   12.4(4)    &    229(8)   &  0.39(17) &    7419.5(90.9) &  1.32(19)  &  0.015(2)   &  $-3.57$(7)E$-06$\\  
ECL-151250 &    0.97177086(46)  &  $ -4.91(10)$E$-6 $  &   26.5(1.0)  &    189(3)   &  0.32(5)  &    6120.8(62.8) &  7.05(22)  &  0.497(24)  &  $-2.16$(8)E$-06$\\  
ECL-161150 &    0.87338659(31)  &  $  -5.19(6)$E$-6 $  &   17.2(5)    &     25(4)   &  0.00(13) &    6401.0(72.5) &  2.64(21)  &  0.062(5)   &  $-6.19$(9)E$-06$\\  
ECL-186358 &    0.44106076(18)  &  $  -3.29(4)$E$-6 $  &   12.6(5)    &    286(12)  &  0.77(13) &    8165.4(79.7) &  1.56(27)  &  0.024(4)   &  $-3.27$(5)E$-06$\\  
ECL-192939 &    0.53189577(18)  &  $  -3.39(4)$E$-6 $  &   14.3(3)    &     49(7)   &  0.63(14) &    9350.8(74.5) &  1.97(20)  &  0.037(4)   &  $-3.26$(4)E$-06$\\  
ECL-212252 &    0.41321226(21)  &  $  -3.91(5)$E$-6 $  &   9.1(6)     &    50(13)   &  0.00(31) &   7071.1(123.6) &  1.13(27)  &  0.018(4)   &  $-4.04$(6)E$-06$\\  
ECL-218355 &    0.44233597(30)  &  $  -4.57(9)$E$-6 $  &   11.9(4)    &    290(7)   &  0.55(12) &    9445.0(84.1) &  2.47(31)  &  0.106(14)  &  $-5.33$(6)E$-06$\\  
ECL-218952 &    0.92756241(41)  &  $ -4.75(13)$E$-6 $  &   12.5(5)    &    275(5)   &  0.00(11) &    8182.2(66.3) &  2.02(17)  &  0.053(5)   &  $-4.90$(8)E$-06$\\  
ECL-220028 &    0.92074865(38)  &  $  -3.84(8)$E$-6 $  &   15.7(7)    &    219(6)   &  0.34(14) &    7758.7(88.4) &  2.12(25)  &  0.039(5)   &  $-4.56$(8)E$-06$\\  
ECL-224851 &    0.37231230(24)  &  $  -4.54(5)$E$-6 $  &   23.0(7)    &      8(2)   &  0.66(5)  &    8155.4(46.9) &  8.68(51)  &  1.234(82)  &  $-2.66$(6)E$-06$\\  
ECL-256434 &    0.45769408(12)  &  $  -3.32(4)$E$-6 $  &   16.0(5)    &    198(9)   &  0.36(34) &   2753.9(120.7) &  0.89(15)  &  0.003(1)   &  $-3.57$(5)E$-06$\\  
ECL-260102 &    0.75241242(11)  &  $  -4.75(3)$E$-6 $  &   12.2(6)    &    233(7)   &  0.11(16) &    7030.6(90.0) &  0.55(07)  &  0.001(1)   &  $-4.46$(5)E$-06$\\  
ECL-264774 &    0.47290437(12)  &  $  -3.90(3)$E$-6 $  &   10.8(2)    &    301(5)   &  0.35(10) &    7901.4(51.1) &  1.54(15)  &  0.031(3)   &  $-3.78$(8)E$-06$\\  
ECL-272254 &    0.85353540(17)  &  $  -5.82(4)$E$-6 $  &   16.7(2)    &    292(2)   &  0.32(04) &    9018.7(32.4) &  2.53(10)  &  0.058(2)   &  $-7.00$(5)E$-06$\\  
ECL-272282 &    0.96391717(45)  &  $ -3.42(12)$E$-6 $  &   19.1(5.7)  &    38(31)   &  0.00(72) &   6678.8(594.9) &  0.43(20)  &  0.001(1)   &  $-3.79$(8)E$-06$\\  
ECL-273301 &    0.44753733(23)  &  $  -3.13(6)$E$-6 $  &   17.2(6)    &    213(4)   &  0.00(12) &    6875.6(73.8) &  3.42(23)  &  0.135(10)  &  $-2.60$(7)E$-06$\\  
ECL-287793 &    0.53189423(12)  &  $  -3.48(3)$E$-6 $  &   12.1(3)    &     71(7)   &  0.22(14) &    9650.9(80.9) &  0.96(11)  &  0.006(1)   &  $-3.23$(6)E$-06$\\  
ECL-288727 &    0.53381448(16)  &  $  -3.18(4)$E$-6 $  &   15.5(1.4)  &    336(7)   &  0.68(12) &    8179.3(88.4) &  1.26(17)  &  0.008(1)   &  $-2.71$(6)E$-06$\\  
ECL-288979 &    0.36994185(17)  &  $  -3.40(4)$E$-6 $  &   11.5(7)    &    210(9)   &  0.00(25) &   6422.7(102.5) &  1.36(23)  &  0.019(4)   &  $-3.23$(6)E$-06$\\  
ECL-289109 &    0.45654460(18)  &  $  -2.99(4)$E$-6 $  &   11.5(1.0)  &    182(17)  &  0.00(63) &   6813.5(201.8) &  0.73(23)  &  0.003(1)   &  $-3.06$(4)E$-06$\\  
ECL-292951 &    0.59568544(14)  &  $  -3.02(4)$E$-6 $  &   9.4(4)     &     17(8)   &  0.21(24) &    5650.6(77.2) &  0.85(12)  &  0.007(1)   &  $-2.77$(6)E$-06$\\  
ECL-293210 &    0.45499924(15)  &  $  -3.47(4)$E$-6 $  &   11.1(3)    &    276(7)   &  0.16(13) &    8431.1(76.3) &  1.33(17)  &  0.019(3)   &  $-3.41$(7)E$-06$\\  
ECL-319624 &    0.60785853(27)  &  $  -3.43(6)$E$-6 $  &   13.1(4)    &    27(10)   &  0.00(30) &   9552.8(132.0) &  1.48(25)  &  0.019(3)   &  $-3.56$(6)E$-06$\\  
ECL-339477 &    1.13814860(38)  &  $  -3.58(8)$E$-6 $  &   17.8(2)    &    340(3)   &  0.08(10) &    8952.2(53.7) &  3.47(21)  &  0.132(8)   &  $-3.42$(14)E$-06$\\ 
\enddata
\end{deluxetable}     
       

\begin{deluxetable}{lccrrrcr}
\tabletypesize{\scriptsize}
\tablewidth{0pt}
\tablecaption{List of Eclipse Timings for 8 CEBs showing a Parabolic Variations}
\tablehead{
\colhead{Object ID}    &\colhead{HJD}     &  \colhead{Error} &\colhead{Cycle} & \colhead{$O-C_1$} & \colhead{$O-C_2$} & \colhead{Source}   &\colhead{$N_{\rm obs}$} \\
\colhead{(OGLE-BLG-)}   &\colhead{(+2450000)}  & \colhead{(day)}  &\colhead{ }     & \colhead{(day)}   & \colhead{(day)}  &       &                      
}       
\startdata       
ECL-069847 & 2454.9030 & 0.0052 & $-6818$ & $-0.17020$ &  $ -0.00031$ & OGLE    &  49    \\
           & 2820.2652 & 0.0034 & $-6270$ & $-0.14578$ &  $ -0.00211$ & OGLE    &  49    \\
           & 5378.4205 & 0.0026 & $-2433$ & $-0.02212$ &  $ -0.00049$ & OGLE    & 126    \\
           & 5726.4325 & 0.0033 & $-1911$ & $-0.01442$ &  $ -0.00107$ & OGLE    & 100    \\
           & 6098.4515 & 0.0037 & $-1353$ & $ 0.00006$ &  $  0.00675$ & OGLE    & 106    \\
           & 6459.7948 & 0.0034 & $ -811$ & $ 0.00556$ &  $  0.00796$ & OGLE    &  89    \\
           & 6841.8006 & 0.0042 & $ -238$ & $ 0.00664$ &  $  0.00685$ & OGLE    &  56    \\
           & 7177.8039 & 0.0072 & $  266$ & $ 0.00574$ &  $  0.00600$ & OGLE    &  64    \\
           & 7858.4587 & 0.0032 & $ 1287$ & $-0.01456$ &  $ -0.00850$ & KMTNet  & 134    \\
           & 7980.4650 & 0.0029 & $ 1470$ & $-0.00973$ &  $ -0.00184$ & KMTNet  & 126    \\
           & 8221.7964 & 0.0053 & $ 1832$ & $-0.01465$ &  $ -0.00239$ & KMTNet  &  82    \\
           & 8225.8157 & 0.0054 & $ 1838$ & $ 0.00463$ &  $  0.01697$ & KMTNet  &  88    \\
\dots      &           &        &         &            &              &         &        \\
\enddata            
\begin{flushleft}
(This table is available in its entirety in machine-readable form.)
\end{flushleft}
\end{deluxetable}        


\begin{deluxetable}{lccrrrrcr}
\tabletypesize{\scriptsize}
\tablewidth{0pt}
\tablecaption{List of Eclipse Timings of 24 CEBs with the parabolic $plus$ LTT ephemeris }
\tablehead{
\colhead{Object ID}    &\colhead{HJD}     &  \colhead{Error} &\colhead{Cycle} & \colhead{$O-C_1$} & \colhead{$O-C_2$} & \colhead{$O-C_3$} & \colhead{Source}   & \colhead{$N_{\rm obs}$} \\
\colhead{(OGLE-BLG-)}   &\colhead{(+2450000)}  & \colhead{(day)}  &\colhead{ }     & \colhead{(day)}   & \colhead{(day)}  & \colhead{(day)} &                    &                      
}       
\startdata       
ECL-088600 & 2448.4717 & 0.0037 & $ -8941.0 $ & $ -0.16670 $ & $  0.00208 $ & $  0.00209 $ & OGLE  &   46    \\   
           & 2821.1195 & 0.0027 & $ -8209.0 $ & $ -0.14760 $ & $ -0.00746 $ & $ -0.00233 $ & OGLE  &   55    \\   
           & 3169.3382 & 0.0022 & $ -7525.0 $ & $ -0.12290 $ & $ -0.00704 $ & $  0.00032 $ & OGLE  &  125    \\   
           & 3541.9905 & 0.0027 & $ -6793.0 $ & $ -0.09920 $ & $ -0.00690 $ & $ -0.00088 $ & OGLE  &  115    \\   
           & 3901.4081 & 0.0025 & $ -6087.0 $ & $ -0.07480 $ & $ -0.00268 $ & $  0.00054 $ & OGLE  &  165    \\   
           & 4268.4583 & 0.0021 & $ -5366.0 $ & $ -0.05350 $ & $  0.00047 $ & $  0.00059 $ & OGLE  &  144    \\   
           & 4631.4313 & 0.0019 & $ -4653.0 $ & $ -0.03720 $ & $  0.00144 $ & $ -0.00124 $ & OGLE  &  166    \\   
           & 4913.4606 & 0.0048 & $ -4099.0 $ & $ -0.02460 $ & $  0.00378 $ & $ -0.00073 $ & OGLE  &   35    \\   
           & 5381.8087 & 0.0029 & $ -3179.0 $ & $ -0.00750 $ & $  0.00725 $ & $  0.00060 $ & OGLE  &  139    \\   
           & 5727.9783 & 0.0021 & $ -2499.0 $ & $  0.00430 $ & $  0.01171 $ & $  0.00437 $ & OGLE  &  242    \\   
           & 6102.6414 & 0.0022 & $ -1763.0 $ & $  0.00250 $ & $  0.00449 $ & $ -0.00251 $ & OGLE  &  291    \\   
           & 6463.5623 & 0.0019 & $ -1054.0 $ & $  0.00310 $ & $  0.00236 $ & $ -0.00298 $ & OGLE  &  263    \\   
\dots      &           &        &         &            &              &         &        \\     
\enddata            
\begin{flushleft}
(This table is available in its entirety in machine-readable form.)
\end{flushleft}
\end{deluxetable}

\end{document}